\documentstyle[12pt]{article}

\newcommand{\BEQ}{\begin{equation}}    
\newcommand{\BEA}{\begin{eqnarray}}
\newcommand{\EEQ}{\end{equation}}      
\newcommand{\EEA}{\end{eqnarray}}
\newcommand{\eps}{\epsilon}                      
\newcommand{\lmb}{\lambda}                       
\newcommand{\sig}{\sigma}                        
\newcommand{\vph}{\varphi}                       
\newcommand{\rar}{\rightarrow}                   
\newcommand{\tia}[1]{\tilde{a}_{#1}}             
\newcommand{\ket}[1]{\left|#1\right\rangle}      
\newcommand{\bra}[1]{\left\langle #1\right|}     
\newcommand{\vm}[1]{\check{#1}}                  
\newcommand{\zeile}[1]{\vskip #1 \baselineskip}  
\newcommand{\vekz}[2]
     {\mbox{${\begin{array}{c} #1  \\ #2 \end{array}}$}}
\newcommand{\matz}[4]
     {\mbox{${\begin{array}{cc} #1 & #2  \\ #3 & #4 \end{array}}$}}

\newcommand{\build}[3]{\mathrel{\mathop{\kern 0pt#1}\limits_{#2}^{#3} }}

\newcommand{\appsection}[2]{\setcounter{equation}{0} \section*{Appendix #1. #2}
\renewcommand{\theequation}{#1.\arabic{equation}}
              \renewcommand{\thesection}{#1} }
\catcode`\@=11
\def\numberbysection{\@addtoreset{equation}{section}
        \def\theequation{\thesection.\arabic{equation}}}
\numberbysection

\begin{document}
%
%
\begin{titlepage}
\null
\begin{center}
{\Large \bf Reaction-Diffusion Processes, Critical
Dynamics and Quantum Chains}
\vskip 0.5in
Francisco C. Alcaraz$^{a}$\footnote{Permanent adress:
Departamento de F\'{\i}sica,
Universidade Federal de S\~{a}o Carlos, 13560 S\~{a}o Carlos SP, Brasil},
Michel Droz$^{b}$, Malte Henkel$^{b}$
and Vladimir Rittenberg$^{a}$
 \\[.3in]
{\em $^{a}$ Physikalisches Institut, Universit\"at Bonn \\
Nu{\ss}allee 12, D - 5300 Bonn 1, Germany}
\zeile{1}
{\em $^{b}$D\'epartement de Physique Th\'{e}orique,
     Universit\'e de Gen\`eve \\
     24  quai Ernest Ansermet,
     CH - 1211 Gen\`eve 4, Switzerland}
\zeile{2}
{\bf UGVA-DPT 1992/12-799}
\end{center}
%
%
\begin{abstract}
The master equation describing non-equilibrium one-dimensional problems
like diffusion limited reactions or critical dynamics of classical
spin systems can be written as a Schr\"odinger equation in which the
wave function is the probability distribution and the Hamiltonian
is that of a quantum chain with nearest neighbor interactions. Since
many one-dimensional quantum chains are integrable, this opens a new field
of applications. At the same time physical intuition and probabilistic
methods bring new insight into the understanding of the properties of
quantum chains. A simple example is the asymmetric diffusion of
several species of particles which leads
naturally to Hecke algebras and $q$-deformed quantum groups.
Many other examples are given.
Several relevant technical aspects like
critical exponents, correlation functions and
finite-size scaling are also discussed in detail.

\end{abstract}

\end{titlepage}

\newpage
%
%

\section{Introduction} \label{sect1}
Our understanding of nonequilibrium statistical physics is far
behind that for the equilibrium theory.
Even simple models may pose a formidable problem
if one wants to approach them analytically.
In this paper, we shall consider two different
types of such problems: the diffusion-limited chemical reactions
and the critical dynamics of classical spin systems.
It will be shown that in one dimension these problems can be
mapped onto quantum chain
problems which are often integrable and
on which a lot of progress was recently
achieved \cite{RMS}. As
a result, many new predictions concerning the
nonequilibrium statistical physics problems follow.

The study of the diffusion-limited chemical reactions
has stimulated a vast amount of
research since the first investigation of Smoluchowski
many years ago \cite{smolo}.
Examples are given by the bimolecular reactions,
$A+B \build{\rightleftharpoons}{g}{k} C+D$ where the two species
$A$ and $B$ diffuse and react to form the two new species $C$ and $D$
and $k$ and $g$ are the forward and backward reaction rates, respectively.
The simple case of irreversible reactions for which $g=0$, $C$ is a inert
product and $D$ is not present has been extensively investigated since the
original work of Zeldowich \cite{zeldo}.
Despite their simplicity those systems
exhibit a very rich dynamical behaviour.
For homogeneous initial conditions and in low dimensions, the
diffusion mechanism is not efficient enough to mix the particles.
As a result a spatial segregation occurs and accordingly,
a reduced number of reactions between the two species is possible.
This results in a slowing down of the
dynamical evolution called  {\sl anomalous kinetics} \cite{kotomin}.
The evolution of the system is not properly described by the usual
rate equations, since the fluctuations play a crucial role.
Another interesting situation is when
the reactants are initially separated in space;
then a reaction-diffusion front is
formed during the evolution \cite{racz}. Here again
the properties of the front are
drastically influenced by the fluctuations in low dimensions \cite{droz}.
We shall restrict ourselves to the homegeneous case here.
Several approaches have been used to study such systems:
numerical simulations \cite{numerique}, scaling arguments
\cite{droz} and analytic arguments based on
the theory of the Brownian motion \cite{Bram91}.
The main results obtained concern the decay
of the number of particles of one species
\cite{decay} and the temporal evolution of the gap developing between
the particles of the two species due to the segregation process \cite{gap}.
However, the analytic results are scarce. Very little is known concerning the
behavior of quantities like the space
and/or time dependent two-particle correlation functions.
Accordingly, new theoretical approches allowing the computation of
such quantities are desirable.

Let us first introduce several models which
have been studied in the literature:
\begin{enumerate}
\item {\it The coagulation model:} \cite{coa} one considers molecules of one
species (say $A$), that diffuse in a milieu and react as:
\BEQ \label{r:coa}
A+A \rightarrow A
\EEQ
\item {\it The annihilation model:} \cite{anni} the $A$
molecules diffuse and annihilate
\BEQ \label{r:anni}
A+A \rightarrow \emptyset
\EEQ
where $\emptyset$ denotes an inert state which decouples completely from
the dynamics.
\item {\it The two-species trapping reaction:} \cite{trapp}
two types of molecules
$A$ and $B$ diffuse and one of them, $B$, is ``trapped'' by $A$:
\BEQ
A+B \rightarrow A
\EEQ
\item {\it The two-species annihilation reaction:} \cite{decay}  two
types of molecules $A$ and $B$ diffuse and annihilate
\BEQ
A+B \rightarrow \emptyset
\EEQ
\end{enumerate}

We are interested in the long-time behaviour of systems like the four
examples given above. Considering quantities like the mean concentration
of $A$ particles $c_{A}(t)$, in general we expect the following two
types of behaviour, as $t\rar\infty$
\BEQ
c_{A}(t) \sim \left\{ \vekz{t^{-\alpha}}{\exp - t/\tau} \right.
\EEQ
where $\alpha$ is some constant and $\tau$ is known as relaxation time.
Throughout this paper, we shall refer to the  first type as ``massless'' or
``critical'', while the second case will
be denoted as ``massive''. This terminology
is borrowed from field theory and
equilibrium statistical mechanics.
All four reactions defined above
have a critical behaviour in the sense that for a given
initial concentration of particles, the long time behaviour of the
concentrations has an algebraic fall-off. For
example, in the reactions (\ref{r:coa}) and (\ref{r:anni}) and in
one-dimension, the concentration $c_A(t)$ of particles behaves  like
\BEQ
c_A(t) \sim t^{-{1}/{2}}
\EEQ
\newline
One can allow for reversible (or back) reactions, corresponding to $g \neq 0$.
In the long time limit ($t \gg 1/g$), a local equilibrium state is reached
\cite{zoltan}.
One is then in a ``massive'' regime, in which the
relaxation towards equilibrium is exponential.

A different type of nonequilibrium problem is the one of
critical dynamics of classical spin systems.
Let us consider for example the case of a classical Ising
model on a lattice.
The system is prepared in an initial (nonequilibrium)
state. One would like to know
how fast the equilibrium state (Gibbs state)
will be reached. When not at (static) criticality,
the system relaxes exponentially towards equilibrium. In general, the
relaxation time $\tau$ scales with
the spatial correlation length $\xi$ as the temperature approaches
its critical value
\BEQ
\tau \sim \xi^z
\EEQ
where $z$ is the dynamical critical exponent.
However, counterexamples will be given in this
paper where $\xi$ diverges but $\tau$ remains finite (no critical
slowing-down!)
As these classical spin systems do not have an
intrinsic dynamics, the dynamics is thought to come from
the interactions between the spins
and a heat bath modelling the fast degrees of freedom
not included in the classical Hamiltonian. The dynamics of
these models is thus formulated in terms of
a master equation for the probability that a spin
configuration is realized at
time $t$ \cite{kawa}. Several cases have to be distinguished
depending on the
presence of macroscopic conserved local quantities.
The simplest case is the purely relaxational one
in which there are no local conservation laws and we shall restrict to this
case throughout the paper. For example, the initial
magnetization relaxes
towards its equilibrium value. All the physics is put
into the transition rates
appearing in the master equation. Several choices are
possible compatible with
the condition of {\sl detailed balance}, which is the condition
insuring that the stationary
state will be the Gibbs one. Despite its apparent simplicity,
the solution of
this problem is very difficult and little can be said
analytically even in
one dimension \cite{felderhof}. The mapping onto
a quantum chain Hamiltonian, which will be
explored throughout the paper, will
turn out to be a very useful
tool to clarify some controversies present in the
literature \cite{cornell}, \cite{julia}.

Both types of nonequilibrium problems (diffusion-limited chemical reactions
 and critical dynamics) can be described in terms of a master
equation for $P(\{\beta\},t)$, the probability that a
configuration $\{\beta\}$
of the systems is realised at time $t$. It turns out that
it is suitable to map this master equation problem onto a
quantum chain problem \cite{felderhof,Glau63,Kada68}.
The corresponding equation of motion reads
\BEA \label{rel1.8}
\partial_t P(\{\beta\},t)=- H P
\EEA
where $H$ is directly related to the transition rates
appearing in the master equation. This will be detailed
in Section~\ref{sect3}. The question of knowing
what kind of dynamical behaviour has the model
(power law or exponential relaxation)
amounts to know in which phase (massive or massless)
of the phase diagram we are.

Moreover, in reaction-diffusion processes $H$ turns out to be non-hermitian
and has often the following particular structure
\BEA \label{rel1.9}
H= H_0+ H_1
\EEA
where $H_0$ is a known integrable Hamiltonian with a larger symmetry
than $H$ (for example, in two-states models it can be the XXZ quantum
chain in a $Z$ field). $H_1$ is non-hermitian and has a lower symmetry but
does not affect the spectrum of $H$. Thus, if
$H_0$ is massless, it follows
that $H$ is massless as well. In problems of critical dynamics, $H$ has
again often the structure of Eq.~(\ref{rel1.9}) but with a new meaning. $H$
is now hermitian, $H_0$ is again a known Hamiltonian (it can be the same
as the one occuring in reaction-diffusion processes) but $H_1$ is now
a perturbation term, if we approach criticality (small temperatures).

When writing this paper we were faced with two problems. The first one
was that we realized that we are left with many more open questions than
answers. This is kind of nice because we hope that this is an invitation
for other people to look closer at the subject. The second problem
concerns the pedagogical presentation of the paper since it addresses two
different communities: people doing nonequilibrium statistical mechanics and
who are familiar with the physical problems
treated in this paper but not with integrable systems and those familiar
with the Bethe ansatz and two-dimensional field theory but not familiar
neither with the physical problems discussed here nor with methods
of computing nonequilibrium averages which are different from those
techniques used in equilibrium statistical mechanics (vacuum expectation
values). We thus suggest
two approaches to this paper. One for the
``mathematician'', the other for the ``physicist''.

The ``mathematician'' should start with Appendix C (the last chapter of the
paper). There we remind the reader
of the definition of the Hecke algebra which depends on a parameter
$\beta=q+q^{-1}$ (the significance of $q$ will become apparent immediately).
As is well known, if a quantum chain can be written as a sum of generators of
a Hecke algebra, through Baxterization \cite{Jone89} one can
associate to the chain an integrable vertex model. In this paper, we will
consider only chains with $2^L$ and $3^L$ states ( $L$ is the length
of the quantum chain). Accordingly, we are going to look for some quotients
of the Hecke algebra. To various quotients of the Hecke algebra one
can associate a representation given by the $(m/n)$ Perk-Schultz
quantum chain \cite{PS}. These are chains with $(m+n)^L$ states invariant
under the quantum superalgebra $U_{q}SU(m/n)$ \cite{Mart92}. In this paper,
the $(2/0)$, $(1/1)$, $(3/0)$ and $(2/1)$ Perk-Schultz models will play
a role with $q$ real
($|\beta|\geq 2$) and the physical significance of
the deformation parameter $q$ will become apparent. We also give some new
representations of the quotients. First
we give non-hermitian representations
(for $q$ real), these are relevant for
expressions like Eq.~(\ref{rel1.9}) and
next, we give representations of the
$(2/0)$ and $(1/1)$ quotients with $3^L$
states. Notice that in the last case
the symmetries of the corresponding chain are
not anymore $U_{q}SU(2/0)$ and $U_{q}SU(1/1)$.
The knowledge of these symmetries
is important because if two chains belong to
the same quotient, their spectrum is
in general the same but the degeneracies are
fixed by the symmetries (a more
detailed version of Appendix C is going to be
published elsewhere). After finishing this
appendix, the ``mathematician'' should go
through Sections~\ref{sect2}-\ref{sect4}
and have a close look at Section~\ref{sectA} where,
in the simple case where the
calculations can be done using free fermions, one illustrates the peculiar
problematics of nonequilibrium statistical mechanics. It is stressed that
(see Eq.~(\ref{rel1.8})) the knowledge of the wave function (not only
the spectrum) plays a crucial role and since in the Bethe ansatz this
knowledge is hard to get, the calculation of average quantities presents
a new challenge. The ``mathematician'' should next have a look a
Appendices A and B and skip Sections~\ref{sect6}-\ref{sect8}.

We sugggest to the ``physicist'' to read the paper
in the chronological order. In Section~\ref{sect2} we consider
quantum chains with $L$ sites (we always take open chains).
On each site we take a discrete variable $\beta$ taking $N$ values
($\beta=0$ corresponds to a vacancy).
We write the most general master equation describing bimolecular reactions.
In Section~\ref{sect3} we write down the corresponding one-dimensional
Hamiltonian, see Eq.~(\ref{rel1.8}). A close related development connecting
the master equation in discrete time to the transfer matrix formalism can
be found in Refs. \cite{Kand90}.

Two-states models are considered in
Section~\ref{sect4}. The elementary processes in the master equation describe
besides diffusion, annihilation, coagulation and death processes (these
processes lead for large times to a state with vacancies only)
also the reverse processes creation, decoagulation and birth. We first show
that for pure diffusion processes which are left-right asymmetric, the
Hamiltonian is just the $q$-deformed XXZ spin-$1/2$ Heisenberg chain.
One discusses in detail the quantum chains corresponding to the different
processes and one stresses the importance of the phase diagram of the
XXZ Heisenberg chain in a Z magnetic field, especially the Pokrovsky-Talapov
line. The learned reader will also notice the importance of non-hermitian
representations of the braid group occuring in this type of problems. As is
well known when a forward-backward process exists (annihilation and
creation for example) through a similarity transformation, the Hamiltonian
is hermitian.
When all three forward-backward processes
are allowed, this is not always possible. We derive the conditions on
the rates in oder to get hermitian Hamiltonians.

Section~\ref{sectA} illustrates
in detail the simple example of annihilation only, with a rate equal to half
the diffusion constant. This corresponds to the physical picture in which,
when two particles are on neighboring sites, they  always annihilate. As
was already known, in this case all the calculations can be done using
free fermions. Here we bring some new results. We consider the
finite-size scaling of the problem (large time and
lattice size with $z=tL^{-2}$ fixed). We show that the finite-size
scaling function exists (no logarithmic corrections).
This result is important because if one accepts its general validity, it
allows numerical estimates of critical exponents like
in equilibrium problems. We also compute, for the first
time, using the Hamiltonian formalism the density-density
correlation function and stress the importance
of the scaling limit $r,t \rar\infty$ with $u=r^2 /t$ fixed.

In Section~\ref{sect5} we consider three-state
models (two species of particles and vacancies)
with $Z_3$ symmetry. Various integrable quantum chains occur
which allows us to obtain some rigrous results. Nevertheless, as will
be seen much work has
still to be done. An interesting three-states model with $Z_2$
symmetry is mentioned in Appendix C. In Section~\ref{sect6} we derive and
solve the condition on the rates in the master equation which gives
chosen steady states. Based on these results, in Sections~\ref{sect7}
we discuss the dynamics of the Ising and
chiral Potts models. This takes us back to the
quantum chains discussed in Sections~\ref{sect4} and \ref{sect5}.
We look at the behaviour of the systems when the temperature is small
(critical dynamics). This brings us to a problem of perturbation theory.
Two cases occur. In the first,
although the (spatial) correlation length diverges,
the relaxation time stays finite.
If we take the corresponding rate to zero and get critical dynamics,
then the relaxation time depends on the (spatial) correlation length
in a way which is independent of the remaining rates (universality).

In Appendix~A we study closer Eq.~(\ref{rel1.9}). We show how from the
knowledge of the eigenvalues and eigenvectors of $H_0$ one can
compute the eigenvectors of $H$. In Appendix B we consider the
example studied in Section~\ref{sectA} from a different point of view.
We notice that the non-hermitian Hermiltonian corresponds to
a representation of the Hecke algebra and using Baxterization, we derive
the corresponding vertex model. It turns out that this is a
seven-vertex model. This observation is relevant, since as
shown in Appendix C (see also \cite{ALVL}) there are other examples
of non-hermitian chains (irreversible processes!)
which satisfy the Hecke algebra and through
this procedure one can find the wave functions using the Bethe ansatz.
Section~10 closes the paper with some open questions.

\section{The master equation} \label{sect2}

In order to write the master equation which describes a general lattice
version of a reaction-diffusion process in
one dimension, we take a chain with
$L$ sites and at each site $i$ we take a variable $\beta_i$ taking $N$ integer
values ($\beta_i = 0,1,\ldots,N-1$).
By convention we attach the value $\beta_i =0$
to a vacancy (inert state). We want to consider a
master equation for the probability distribution
$P(\{\beta\};t)$ with the following form
\BEA
\lefteqn{ \frac{\partial P(\{\beta \};t)}{\partial t} =
\sum_{k=1}^{L-1} \left[ - w_{0,0}^{(k)} (\beta_k ,\beta_{k+1})
P(\beta_1,\ldots ,\beta_L ;t) \vekz{ }{~} \right. } \nonumber \\
& & + \left. {\mathop{{\sum}'}_{\ell,m=0}^{N-1}}
w_{\ell,m}^{(k)}(\beta_k,\beta_{k+1}) P(\beta_1,\ldots,\beta_k +\ell,
\beta_{k+1}+m,\ldots,\beta_L ;t) \right] \label{eq:2.1}
\EEA
where $w_{\ell,m}^{(k)}$ are the transition
rates and the prime in the second sum
indicates that the pair $\ell =m=0$ should be excluded.
We assume all the additions performed on the $\beta_i$ to be done modulo $N$.

The advantage of this notation is that one
can introduce discrete symmetries in
a simple way. We shall assume hereafter that
the system is homogenous which implies that
the transitions rates are independent of $k$
\BEQ
w_{\ell,m}^{(k)}(\alpha,\beta) = w_{\ell,m}(\alpha,\beta)
\EEQ
for all $k=1,\ldots,L-1$.
The probability $\Gamma_{(\alpha,\beta)}^{(\gamma,\delta)}$ that a state
$(\gamma,\delta)$ on two consecutive sites
will change after an unit time into the state $(\alpha,\beta)$
is
\BEQ \label{transrat}
\Gamma_{(\alpha,\beta)}^{(\gamma,\delta)}=
w_{\gamma-\alpha,\delta-\beta}(\alpha,\beta) \;\; ; \;\;
(\alpha,\beta) \neq (\gamma,\delta)
\EEQ
The rates $w_{0,0}(\alpha,\beta)$ are related to the probability that in
the unit time the state $(\alpha,\beta)$  unchanges.
{}From the conservation of probabilities, we have
\BEQ \label{gl2.5}
w_{0,0}(\alpha,\beta) = {\sum_{r,s}}' w_{r,s}(\alpha-r,\beta-s)
\EEQ
where $r=s=0$ is again excluded.

It is now trivial to check, using
Eq.~(\ref{transrat}), that if for the $N$-state model we want to have a $Z_N$
symmetry, only the functions
\BEQ \label{gl2.6}
w_{\ell,m}(\alpha,\beta) \;\; , \;\; \ell + m = 0 (\mbox{\rm modulo } N)
\EEQ
will appear.
This is the case of the annihilation model
( written as $A+A\rar\emptyset + \emptyset$ where
$\emptyset$ is a vacancy) which has $Z_2$
symmetry ($N=2$) and for the two-species
annihilation reaction ($N=3$ and the reaction is written as
$A+B\rar\emptyset + \emptyset$) which has $Z_3$ symmetry. In
the former case, we assign to the vacancy the $Z_2$ quantum number $0$ and
to the $A$ molecule the number $1$. In the latter case, the vacancy is
denoted by $0$, the $A$ by $1$ and the $B$ by $2$.

Parity conservation (left-right symmetry) is achieved if
\BEQ \label{rel2.6}
w_{\ell,m}(\alpha,\beta) = w_{m,\ell}(\beta,\alpha)
\EEQ

Let us comment on the supplementary symmetries
besides a possible parity invariance which exists in the four
examples given in Section~\ref{sect1}.
The coagulation model has no symmetry at all.
As we have seen, the annihilation model has
a $Z_2$ symmetry. The two-species trapping reaction has a $U(1)$ symmetry
(since the number of $A$ particles is conserved) and a $Z_2$ symmetry
(give to the vacany and the state $B$ a $Z_2$-parity ``+'' and to
the state $A$ a $Z_2$-parity ``-'').
As mentioned before, the two-particle reaction has
a $Z_3$ symmetry. It has also a supplementary $U(1)$ symmetry, since the
difference of the numbers of $A$ and $B$ particles is conserved.

It is often useful to make a change of variables in the master equation.
In this paper, we define
\BEQ \label{gl2.8}
P(\{\beta\};t) = \Phi(\{\beta\}) \Psi(\{\beta\};t)
\EEQ
where $\Phi(\{\beta\})$ takes the special form
\BEQ \label{eq:2.8}
\Phi(\{\beta\}) = \prod_{k=1}^{L} h^{(k)}(\beta_k)
\EEQ
{}From Eq.~(\ref{eq:2.1}) $\Psi$ is a solution of the new master equation
\BEA
\frac{\partial}{\partial t}\Psi(\{\beta\};t) &=&
\sum_{k=1}^{L-1} \left[ -w_{0,0}(\beta_k,\beta_{k+1}) \Psi(\{\beta\};t)
\vekz{ }{~} \right. \nonumber \\
& & + \left. {\sum_{\ell,m}}'
W^{(k)}_{\ell,m}(\beta_k,\beta_{k+1})
\Psi (\beta_1,\ldots,\beta_k +\ell, \beta_{k+1}+m,\ldots,\beta_{L};t)
\right] \nonumber \\ & & \label{gl2.10}
\EEA
with the pair $\ell=m=0$ excluded and
\BEQ \label{gl2.11}
W^{(k)}_{\ell,m}(\alpha,\beta) = \frac{\vph^{(k)} (\alpha+\ell,\beta+m)}
{\vph^{(k)}(\alpha,\beta)} w_{\ell,m}(\alpha,\beta)
\EEQ
where
\BEQ \label{alela}
{\vph^{(k)} (\alpha,\beta)} =
 {h^{(k)}(\alpha) h^{(k+1)}(\beta)} \;\; ; \;\; k = 1, 2, ..., L-1.
\EEQ
Notice that although the rates $w_{\ell,m}(\alpha,\beta)$ are link independent
in general, the $W^{(k)}_{\ell,m}(\alpha,\beta)$ are not.
The function $\Phi(\{\beta\})$ has obviously to be non-zero and finite.
Let us finally note that if we want all ``molecules'' to
disappear at large times so that
we are only left with vacancies, we must have the condition
\BEQ \label{gl2.12}
w_{\ell,m}({-\ell,-m})= 0
\EEQ
for $\ell,m =0,1,\ldots,N-1$. One can check that
\BEQ \label{gl2.13}
P(\{\beta\};t) = \prod_{k=1}^{L-1} \delta(\beta_k)
\EEQ
satisfies indeed $\partial P/\partial t = 0$, with the convention that
$\beta_k=0$ corresponds to a vacancy.

\section{Quantum chains corresponding to master equations}\label{sect3}

We shall now write the master equation ~(\ref{gl2.10}) in the form
of a Schr\"odinger equation. In order to do so, on each site we define
a basis in the space of $N\times N$ matrices $E^{k\ell}$. The only
non-vanishing matrix element of the matrix $E^{k\ell}$ is the one
on the $k^{th}$ line and the $\ell^{th}$ column and this matrix element
is equal to unity. Assuming homogeneity, the most general $N^{L} \times
N^{L}$ Hamiltonian with only nearest-neighbor interactions  can be written as
\BEQ \label{rel3.1}
H = \sum_{i=1}^{L-1} H_i
\EEQ
where
\BEQ
H_i = \sum a_{\ell,m,r,s} E^{\ell m} \otimes E^{rs}
\EEQ
acts in the subspace $V^{(i)} \otimes V^{(i+1)}$ of the
$N^{L}$-dimensional vector
space. Besides the matrices $E^{k\ell}$, on each site it is also convenient
to define the matrix $F$
\BEQ
F = \left( \begin{array}{cccccc}
0 & 1 & 0 & 0 & \cdots & 0 \\
0 & 0 & 1 & 0 & \cdots & 0 \\
\vdots  &   &   & \ddots &   & \vdots  \\
\vdots  &   &   &    & \ddots & \vdots \\
0 & 0 & 0 & \cdots & 0 & 1 \\
1 & 0 & 0 & \cdots & 0 & 0 \end{array} \right) \;\; ; \;\;
F^{N} = 1
\EEQ
With these notations, the Hamiltonian density can be written as
\BEQ \label{gl3.4}
H_i = U_i - T_i
\EEQ
where
\BEA
T_i &=& {\mathop{{\sum}'}_{\ell,m=0}^{N-1}} \sum_{\alpha,\beta=0}^{N-1}
W^{(i)}_{\ell,m}(\alpha,\beta) \left( E^{\alpha \alpha} F^{\ell}\right) \otimes
\left( E^{\beta \beta} F^m \right) \label{gl3.5a} \\
U_i &=& \sum_{\alpha,\beta=0}^{N-1} w_{0,0}(\alpha,\beta)
E^{\alpha \alpha} \otimes E^{\beta \beta} \label{gl3.5b}
\EEA
Here $W^{(i)}_{\ell,m}$ is given in Eq.~(\ref{gl2.11}) and $w_{0,0}$ in
Eq.~(\ref{gl2.5}). The Schr\"odinger equation replacing the master equation
reads
\BEQ \label{rel3.7}
\frac{\partial}{\partial t} \ket{\Psi}= - {H} \ket{\Psi}
\EEQ
The Schr\"odinger equation corresponding to the physical problem
(without the similarity transformation Eq.~(\ref{gl2.8})) is
\BEQ \label{rel3.8}
\frac{\partial}{\partial t} \ket{P}= - \widetilde{H} \ket{P}
\EEQ
$\widetilde{H}$ is obtained by taking $w_{\ell,m}(\alpha,\beta)$ instead of
$W_{\ell,m}(\alpha,\beta)$ in Eq.~(\ref{gl3.5a}). Obviously $H$ and
$\widetilde{H}$ have the same spectra
but they are given by two different one-dimensional quantum chains
which act in a $N^L$ Fock space (for $N-1$ species)
and their explicit form depends on the chemical reaction and diffusion
process. We first discuss the Schr\"odinger equation (\ref{rel3.8})
The ket state $\ket{P}$ is defined as follows \cite{Kada68}. Take an orthogonal
basis in $\{\beta\}$
\BEQ
\ket{\{\beta\}} = \ket{\beta_1, \ldots, \beta_L} \;\; ; \;\;
\bra{\{\beta'\}} \ket{\{\beta\}} = \delta_{ \{\beta'\},\{\beta\}}
\EEQ
then
\BEQ
\ket{P} = \sum_{\{\beta\}} P(\{\beta\};t) \ket{\{\beta\}}
\EEQ
The reaction-diffusion process is determined by the
initial ($t=0$) probability distribution $P_{0}(\{\beta\})$
which defines the ``initial'' ket state
\BEQ
\ket{P_0} = \sum_{\{\beta\}} P_{0}(\{\beta\}) \ket{\{\beta\}}
\EEQ
The Hamiltonian $\widetilde{H}$ is in general non-hermitian and
due to probability
conservation (Eq.~(\ref{gl2.5})), it satisfies the remarkable relation:
\BEQ \label{rel3.12}
\bra{0} \widetilde{H} = 0
\EEQ
where the bra ground state $\bra{0}$ is
\BEQ
\bra{0} = \sum_{\{\beta\}} \bra{\{\beta\}}
\EEQ
{}From Eq.~(\ref{rel3.12}) it follows that the ground state energy is zero.
Take now an observable $X(\{\beta\})$
(for example the concentration of $A$
particles in the coagulation model (\ref{r:coa})). Its average can be
computed as follows
\BEA
<X>(t) &=& \sum_{\{\beta\}} X(\{\beta\}) P(\{\beta\};t) \nonumber \\
&=& \bra{0} X \ket{P} =
\bra{0} X e^{-\widetilde{H} t} \ket{P_0} \label{rel3.14}
\EEA
Notice that that in nonequilibrium problems one studies the properties of
the wave function which is already a probability and {\em not}
quantum mechanical expectation values
\BEQ
\bra{0} X \ket{0}
\EEQ
as one does in equilibrium problems. The thermodynamical (continuum) limit
can be computed from Eq.~(\ref{rel3.14}), taking the length $L$ of the chain
to infinity for a fixed time $t$. As discussed in detail in
Section~\ref{sectA},
a second limit (the
finite-size scaling limit) is also interesting, where one
takes both $t$ and $L$ large but keeps $z = t/L^2$ finite.

If ${E}_{\lmb}$ and $\ket{\Psi_{\lmb}}$ are
the eigenvalues and eigenkets of $\widetilde{H}$,
we have from Eq.~(\ref{rel3.14})
\BEQ
<X>(t) = \sum_{\lmb} a_{\lmb} e^{-{E}_{\lmb}t}
\bra{0}X\ket{\Psi_{\lmb}}
\EEQ
where
\BEQ
\ket{P_0} = \sum_{\lmb} a_{\lmb} \ket{\Psi_{\lmb}}
\EEQ
Thus the large $t$ behaviour of $<X>(t)$ is governed by the lowest excitations
of $\widetilde{H}$. If, instead of
$\widetilde{H}$ we use $H$ (see Eq.~(\ref{rel3.7})), then
the averages have different expressions
\BEQ \label{rel3.18}
<X>(t) = \sum_{\{\beta\}} X(\{\beta\}) \Phi(\{\beta\}) \Psi(\{\beta\};t)
= \bra{0} X \Phi e^{-Ht}\ket{\Psi_0}
\EEQ
where
\BEQ
\ket{\Psi_0} =
\sum_{\{\beta\}} \Phi^{-1}(\{\beta\}) P_{0}(\{\beta\}) \ket{\{\beta\}}
\EEQ

There is another distinctive feature of nonequilibrium problems, as compared to
equilibrium ones, and this is the concept of interaction range. As opposed to
equilibrium problems where the whole information is only contained in the
Hamiltonian (it is nearest-neighbour interactions as can be seen from
Eq.~(\ref{rel3.1})),
in the nonequilibrium case we have
to give also $P_{0}(\{\beta\})$. This probability can
describe an uncorrelated homogenuous
distribution like
\BEQ
P_{0}(\{\beta\}) = \prod_{i=1}^{L} f(\beta_i)
\EEQ
or a strongly correlated distribution when, for example, at $t=0$ the reactants
are separated in space \cite{droz}. The general properties of
$P(\{\beta\};t)$ and implicitly those of averages (like self-organization,
critical dimensions or critical exponents) are going to be different. This
can be understood when comparing Eqs.~(\ref{rel3.14}) and (\ref{rel3.18}). Let
us assume that we give $\widetilde{H}$ and $P_0$
describing a correlated distribution,
we can make a similarity transformation
to bring $P_0$ to an uncorrelated distribution
$\Psi_0$. After this transformation we will have to work with a new Hamiltonian
$H$ (in general with long-range interactions)
with different physical properties.

\section{Two-state Hamiltonians} \label{sect4}

Since for two-states models instead of the basis $E^{\alpha,\beta}$ one often
prefers the basis of Pauli matrices, let us start by giving some
useful identities
\BEA
E^{01}\otimes E^{10} + E^{10}\otimes E^{01} &=& \frac{1}{2}
\left( \sig^{x} \otimes \sig^{x} + \sig^{y} \otimes \sig^{y} \right)
\nonumber \\
E^{01} \otimes E^{01} &=& \frac{1}{4} \left( \sig^{x} \otimes \sig^{x}
- \sig^{y} \otimes \sig^{y} + i \left( \sig^{x} \otimes \sig^{y}
+ \sig^{y} \otimes \sig^{x} \right) \right) \nonumber \\
E^{01} \otimes E^{01} + E^{10} \otimes E^{10} &=&
\frac{1}{2} \left( \sig^{x} \otimes \sig^{x} - \sig^{y} \otimes \sig^{y}
\right)
\EEA
We are now going to consider various Hamiltonians according to their
symmetries and chemical properties. In the two-states models, we
have $A$ states and vacancies.

\subsection{$Z_2$ symmetric, parity non-invariant vacuum-driven processes}

By a vacuum-driven process, or inert-driven process,
we mean reactions which end in a state with
vacancies only. This implies from Eq.~(\ref{gl2.12})
\BEQ \label{rel4.2}
w_{1,0}(1,0) = w_{0,1}(0,1) = w_{1,1}(1,1) = 0
\EEQ
This means that there is no production of pairs of $A$ particles, in other
words the process $\emptyset + \emptyset \rar A + A$ is forbidden. Since
\BEQ
w_{1,0}(\alpha,\beta) = w_{0,1}(\alpha,\beta) = 0
\EEQ
because of the $Z_2$ symmetry, we are left only with the processes
\begin{enumerate}
\item[a.)] annihilation, with the rate $w_{1,1}(0,0)$
\BEQ
A + A \rar \emptyset + \emptyset
\EEQ
\item[b.)] diffusion to the right, with the rate $w_{1,1}(0,1)$
\BEQ
A + \emptyset \rar \emptyset + A
\EEQ
\item[c.)] diffusion to the left, with the rate $w_{1,1}(1,0)$
\BEQ
\emptyset + A \rar A + \emptyset
\EEQ
\end{enumerate}
We use now Eqs.~(\ref{gl3.4}, \ref{gl3.5a}, \ref{gl3.5b}), take
into account that
\BEQ
E^{00} F = E^{01} \;\; , \;\; E^{11} F = E^{10} \;\; , \;\;
\sig^{z} = E^{00} - E^{11},
\EEQ
and find
\BEA
T_{i} &=& \frac{\vph^{(i)}(1,1)}{\vph^{(i)}(0,0)} w_{1,1}(0,0) E^{01} \otimes
E^{01} \nonumber \\
& & + \frac{\vph^{(i)}(0,1)}{\vph^{(i)}(1,0)} w_{1,1}(1,0) E^{10} \otimes
E^{01} + \frac{\vph^{(i)}(1,0)}{\vph^{(i)}(0,1)} w_{1,1}(0,1) E^{01} \otimes
E^{10} \nonumber \\
U_{i} &=& \frac{1}{4} \left( w_{1,1}(0,1) + w_{1,1}(1,0) +
w_{1,1}(0,0) \right) {\bf 1} \otimes {\bf 1} \nonumber \\
& & + \frac{1}{4} \left( w_{1,1}(0,0) - w_{1,1}(0,1) - w_{1,1}(1,0)
\right) \sig^{z} \otimes \sig^{z} \nonumber \\
& & - \frac{1}{4} w_{1,1}(0,0) \left( \sig^{z} \otimes {\bf 1} +
{\bf 1} \otimes \sig^{z} \right) \nonumber \\
& & + \frac{1}{4} \left( w_{1,1}(1,0) - w_{1,1}(0,1) \right)
\left( \sig^{z} \otimes {\bf 1} - {\bf 1} \otimes \sig^{z} \right)
\label{eq:4.8}
\EEA
Up to this point we still have freedom in the choice of the function
$\vph^{(k)}(\beta_{k},\beta_{k+1})$ in Eq.~(\ref{eq:2.8}).
We now take
\BEQ \label{eq:4.9}
\frac{\vph^{(k)}(1,0)}{\vph^{(k)}(0,1)} =
\sqrt{ \frac{w_{1,1}(1,0)}{w_{1,1}(0,1)} } = q \;\; ; \;\; k = 1, 2, ..., L-1,
\EEQ
and it is clear that $q$ is real since the rates are real. From
Eqs.~(\ref{eq:4.9}) and (\ref{alela}) we get
\BEQ \label{fii}
\frac{h^{(k)}(1)}{h^{(k)}(0)} = q^{1-k}\lambda^{-1},
\EEQ
where $\lambda$ is an arbitrary parameter.
The diffusion constant $D$, expressed as
\BEQ \label{eq:4.10}
D = \sqrt{w_{1,1}(0,1) w_{1,1}(1,0)},
\EEQ
fixes the time scale. We then finally obtain
\BEQ \label{gl4.8}
H = D \left( H_0 + H_1 \right)
\EEQ
where
\BEA
H_0 &=& - \frac{1}{2} \sum_{i=1}^{L-1} \left[ \sig_{i}^{x} \sig_{i+1}^{x}
+ \sig_{i}^{y} \sig_{i+1}^{y} + \Delta \sig_{i}^{z} \sig_{i+1}^{z}
+ (1 - \Delta')\left( \sig_{i}^{z} + \sig_{i+1}^{z} \right) \right.
\nonumber \\
 & & \left. - \frac{1}{2} \left( q - q^{-1} \right) \left(
\sig_{i}^{z} - \sig_{i+1}^{z} \right) + 2 \Delta' - \Delta -2
\right] \label{gl4.9a} \\
H_1 &=& - \frac{w_{1,1}(0,0)}{\lambda^2 D}
\sum_{i=1}^{L-1} q^{1-2i}E_{i}^{01} E_{i+1}^{01} \label{gl4.9c}
\EEA
\BEQ
\Delta = \frac{q+q^{-1}}{2} - \frac{w_{1,1}(0,0)}{2D} \;\; , \;\;
\Delta' = 1 - \frac{w_{1,1}(0,0)}{2D} \label{gl4.9b}
\EEQ
Let us now discuss the structure of this Hamiltonian. $H_0$ is hermitian,
$U(1)$ symmetric and its properties are going to be discussed
shortly. $H_1$ is non-hermitian and has only a $Z_2$ symmetry, corresponding
to the transformations
\BEQ
\sig_{i}^{x} \rar - \sig_{i}^{x} \;\; , \;\;
\sig_{i}^{y} \rar - \sig_{i}^{y} \;\; , \;\;
\sig_{i}^{z} \rar  \sig_{i}^{z}
\EEQ
which makes $H$ only $Z_2$-symmetric (as expected) and non-hermitian.

Let us discuss the case of pure diffusion, that is $w_{1,1}(0,0)=0$.
We have $H_1 =0$ and
\BEQ
H_0 = - \frac{1}{2} \sum_{i=1}^{L-1} \left[ \sig_{i}^{x} \sig_{i+1}^{x}
+ \sig_{i}^{y} \sig_{i+1}^{y} + \frac{q+q^{-1}}{2}
\sig_{i}^{z} \sig_{i+1}^{z}  -  \frac{q - q^{-1}}{2} \left(
\sig_{i}^{z} - \sig_{i+1}^{z} \right) - \frac{q+q^{-1}}{2}
\right] \label{eq:purediff}
\EEQ
This Hamiltonian is the $U_{q}SU(2)$ symmetric Hamiltonian of
Pasquier and Saleur \cite{Pasq90}
where the deformation parameter $q$ has a clear physical interpretation, see
Eq.~(\ref{eq:4.9}). Moreover, as shown in Appendix C, $H_0$ is the sum of
generators of a given quotient of a Hecke algebra (called a
Temperley-Lieb algebra). As discussed in detail in
Appendix C, for any number of species, asymmetric diffusion defines various
quotients of Hecke algebras (we think this is an important observation).
As a corollary one can show that for any number of
species asymmetric diffusion is always massive.

We now consider the case $w_{1,1}(0,0) \neq 0$, that is the full nonhermitian
Hamiltonian Eq.~(\ref{gl4.9a}) which has a special property. From the
expression ~(\ref{gl4.9b}) of $H_1$ we see that one can take
the factor $\lambda^2$
arbitrary without changing the spectrum. This is certainly so, since this
factor only serves to parametrize a similarity transformation. On the
other hand, one can also see by direct inspection of the matrix elements of $H$
that the characteristic polynomial of $H$ is not affected by the
existence of $H_1$. Let us briefly present the argument since we are going
to use this repeatedly later on. If $C$ and $D$ are square matrices and if $X$
is a rectangular matrix, it is well known that
\BEQ \label{eq:4.16}
\det \left( \matz{C}{X}{0}{D} \right) = \det C \det D
\EEQ
Now, $H_0$ has a $U(1)$ symmetry and has thus a block-diagonal form.
The corresponding quantum number labelling the blocks is the
number of $A$ particles present. Acting with $H_1$ reduces this quantum number
by two and this plays the role of $X$ in Eq.~(\ref{eq:4.16}).
Thus the above determinant formula can be applied and the
independence of the spectrum of $H$ from the operator $H_1$ follows.

Let us choose $w_{1,1}(0,0)$ such that $\Delta=0$.
This corresponds accidentally to the
choice of all the previous numerical simulation studies
of the one-species annihilation
process where one took $q=1$ and $w_{1,1}(0,0)=2D$.
In this case $H$ can be studied in terms of free fermions.
This will be done in Section~\ref{sectA} where some new results
are presented. In particular we find that for
$q=1$, $H_0$ can be written
in terms of fermionic number operators as follows
\BEQ \label{gl4.12}
H_0 \sim \sum_k \left( \frac{k}{L} \right)^{2} a_{k}^{+} a_{k}
\EEQ
which describes a Pokrovsky-Talapov phase transition \cite{PT}. Notice the
quadratic dispersion relation which is expected in any diffusion
problem.

In Appendix~B we consider a complementary point of view
which is related to the problem of integrability of quantum chains
with the structure Eq.~(\ref{gl4.8}) in which $H_0$ is integrable
(in the present case it corresponds to a six-vertex model)
and $H_1$ does not affect the spectrum of $H$.
We show that $H$ is related to a seven-vertex
model for which the Yang-Baxter equations
are valid and hence $H$ is integrable.

We now consider the case $\Delta\neq 0$.
This brings us, see Eq.~(\ref{gl4.9a}),
to discuss properties of the Hamiltonian
\BEQ \label{ariro}
H' = -\frac{1}{2} \sum_{i=1}^{L-1} \left[ \sig_{i}^{x} \sig_{i+1}^{x}
+ \sig_{i}^{y} \sig_{i+1}^{y} + \Delta \sig_{i}^{z} \sig_{i+1}^{z}
+ h \left( \sig_{i}^{z} + \sig_{i+1}^{z}\right)  \right]
\EEQ
This Hamiltonian is
integrable \cite{Mcco73} and its phase diagram is shown in Fig.~1.
For $h=0$, the system is massive with a ferromagnetic ground state
if $\Delta > 1$, massless and conformally invariant if
$-1 \leq \Delta \leq 1$ and again massive with an antiferromagnetic ground
state if $\Delta < -1$. For a given $h$, the system is massive
ferromagnetic if $\Delta > 1 - h$, then undergoes a
Pokrovsky-Talapov transition at $\Delta_{PT} = 1 - h$, is in a massless
incommensurate phase for $\Delta_{c} < \Delta < 1 - h$
and reaches again the massive antiferromagnetic phase if
$\Delta < \Delta_c$.

It can be shown \cite{zuzeigen} that in the continuum, along
the Pokrovsky-Talapov line, the spectrum of $H'$ is, up to
normalisation, given by Eq.~(\ref{gl4.12}) for any $\Delta$.
In other words, the system is massless
with a quadratic dispersion in the momentum $k$.

With this knowledge in hand, let us discuss some properties of $H_0$.
Since $\Delta' < \Delta$, we have $h > 1 - \Delta$ and the system
is massive. If however, $q=1$, we have $\Delta'=\Delta$ and we are
on the Pokrovsky-Talapov line where the system is massless. There are
some immediate questions to ask about this system. If we define
\BEQ
\eps = \frac{q+q^{-1}}{2} -1
\EEQ
and if $\tau$ denotes the relaxation time, we are interested in the
exponent $\eta$
\BEQ \label{eq:4.20}
\tau \sim \eps^{-\eta}
\EEQ
 It can be shown that
$\eta = 1$ is independent of $\Delta$ using the standard lore of Hecke
algebras,
see Appendix~C. This result is to be expected since we can change the
value of $\Delta$ by changing $D$, see Eq.~(\ref{gl4.9b}). However, changing
$D$ merely changes the time scale.
More generally, the full finite-size scaling form should read
\BEQ \label{rel4.21}
\tau = L^{\eta_1} F(x) \;\; , \;\; x = \eps L^{\eta_2}
\EEQ
in the simultaneous limit $\eps \rar 0, L \rar \infty$ with $x$
kept fixed. Concerning the concentration of $A$ particles
\BEQ
c_{A} = \frac{1}{L} \sum_{i=1}^{L} E_{i}^{11}
\EEQ
in the critical regime ($q=1$), we are interested in
\BEQ \label{rel4.23}
<c_A >(t) = L^{x} \phi (z) \;\; , \;\; z = t L^{-2}
\EEQ
for both $t$ and $L$ being large with $z$ fixed. The exponent
$2$ in the definition of $z$ was taken because of the quadratic
dispersion relation
Eq.~(\ref{gl4.12}). In Section~\ref{sectA} we shall find $x = -1$.
Similarly, correlation functions can be introduced
and calculated. We shall present the analysis in Section~\ref{sectA}
for the case $\Delta=0$. For the case $\Delta \neq 0$, although we expect on
physical grounds the same results, the explicit proof is still
missing. The Bethe ansatz equations for the wave equations are known only
for $H_0$ but not for $H$. In Appendix~A it is shown how in
principle the knowledge of the eigenfunctions of $H_0$ can help to find the
eigenfunctions of $H$. In a different approach one could start with the
seven-vertex model (which is not limited to
the case $\Delta =0$), and perform
the Bethe ansatz there. The wave functions thus found could be used for $H$.
Whether this whole program is manageable remains to be seen.

\subsection{$Z_2$ non-invariant, parity invariant vacuum-driven processes}

{}From now on, we take always the left-right symmetric case, see
Eq.~(\ref{rel2.6}), in all
reaction rates, in particular $w_{1,1}(0,1)=w_{1,1}(1,0)=D$. We choose
units of time such that $D=1$. To
the processes studied before we now add the following
\begin{enumerate}
\item[d.)] coagulation, with rate $w_{1,0}(0,1)$
\BEQ
A + A \rar \emptyset + A
\EEQ
\item[e.)] decoagulation, with rate $w_{1,0}(1,1)$
\BEQ
\emptyset + A \rar A + A
\EEQ
\item[f.)] death, with rate $w_{1,0}(0,0)$
\BEQ
A + \emptyset \rar \emptyset + \emptyset
\EEQ
\end{enumerate}
We take $\vph^{(i)}(0,1)=\vph^{(i)}(1,0) (i = 1, 2, ..., L-1)$ and get
\BEA
T_i &=&  \left( E^{01} \otimes E^{10} + E^{10} \otimes E^{01}
\right) + \frac{\vph^{(i)}(1,1)}{\vph^{(i)}(0,0)} {w_{1,1}(0,0)}
E^{01} \otimes E^{01} \nonumber \\
& & + \frac{\vph^{(i)}(1,0)}{\vph^{(i)}(0,0)} {w_{1,0}(0,0)}
\left( E^{01} \otimes E^{00} + E^{00} \otimes E^{01} \right) \nonumber \\
& & + \frac{\vph^{(i)}(1,1)}{\vph^{(i)}(1,0)} {w_{1,0}(0,1)}
\left( E^{01} \otimes E^{11} + E^{11} \otimes E^{01} \right) \nonumber \\
& & + \frac{\vph^{(i)}(1,0)}{\vph^{(i)}(1,1)} {w_{1,0}(1,1)}
\left( E^{10} \otimes E^{11} + E^{11} \otimes E^{10} \right) \label{gl4.19a} \\
U_i &=& \frac{1}{2}{w_{0,0}(1,0)}
+ \frac{1}{4} {w_{0,0}(1,1)}
+ \left( \frac{w_{0,0}(1,1)}{4} - \frac{w_{0,0}(1,0)}{2}
\right) \sig^{z} \otimes \sig^{z} \nonumber \\
& & - \frac{w_{0,0}(1,1)}{4} \left( {\bf 1} \otimes \sig^{z}
+ \sig^{z} \otimes {\bf 1} \right)
\EEA
where
\BEA
w_{0,0}(1,1) &=& 2 w_{1,0}(0,1) + w_{1,1}(0,0) \nonumber \\
w_{0,0}(1,0) &=& w_{1,0}(0,0) + w_{1,0}(1,1) + 1
\EEA
Thus the Hamiltonian depends on four parameters. Two cases have to
be considered separately. In the first case we have only coagulation,
in the second case we have coagulation and decoagulation. If we have no
decoagulation, $w_{1,0}(1,1)=0$. Then
\BEQ
H = H_0 + H_1
\EEQ
where
\BEA
H_0 &=& -\frac{1}{2} \sum_{i=1}^{L-1} \left[ \sig_{i}^{x} \sig_{i+1}^{x}
+ \sig_{i}^{y} \sig_{i+1}^{y} + \Delta \sig_{i}^{z} \sig_{i+1}^{z}
+ (1 - \Delta') \left( \sig_{i}^{z} + \sig_{i+1}^{z} \right)
+2\Delta' - \Delta -2 \right] \nonumber \\ & & \label{gl4.21} \\
H_1 &=& - \sum_{i=1}^{L-1} \left[ \frac{\vph^{(i)}(1,1)}{\vph^{(i)}(0,0)}
{w_{1,1}(0,0)} E_{i}^{01} E_{i+1}^{01}
+ \frac{\vph^{(i)}(1,0)}{\vph^{(i)}(0,0)} {w_{1,0}(0,0)}
\left( E_{i}^{01} E_{i+1}^{00} + E_{i}^{00} E_{i+1}^{01}
\right) \right. \nonumber \\
& & + \left. \frac{\vph^{(i)}(1,1)}{\vph^{(i)}(1,0)}
{w_{1,0}(0,1)} \left( E_{i}^{01} E_{i+1}^{11}
+ E_{i}^{11} E_{i+1}^{01} \right) \right]
\EEA
\BEA
\Delta' &=& 1 - {w_{1,0}(0,1)} - \frac{1}{2}
{w_{1,1}(0,0)} \\
\Delta &=& \Delta' + {w_{1,0}(0,0)}
\EEA
We have checked using the same arguments as above that
the spectrum of $H$ is independent of the presence of $H_1$. Since $\Delta'
< \Delta$, we are in a massive phase. The system is massless, i.e.
$\Delta=\Delta'$, if $w_{1,0}(0,0) =0$, corresponding to the absence of
death processes. A straightforward calculation shows that the relaxation time
$\tau$ scales like
\BEQ \label{eq:tauscal}
\tau \sim \left( \frac{w_{1,0}(0,1)}{w_{1,1}(1,0)} \right)^{-1}
\EEQ
This can be seen by repeating the same arguments as for Eq.~(\ref{eq:4.20}).
The result Eq.~(\ref{eq:tauscal}) can be understood easily if one keeps
in mind the time evolution of the system. If one considers the later
stages of the process when few ``molecules'' are left, few of them will
meet and annihilate. On the other hand, any one of them can ``die'' since this
is an individual process.

In the absence of the death process some results are already known for the
pure annihilation process \cite{anni} or the coagulation process \cite{coa}
and it remains to be seen how much more can be learnt using the
quantum chain formulation.

We now consider the case when the coagulation and the decoagulation
processes coexist and we have all terms in Eq.~(\ref{gl4.19a}). Then little
can be said. One interesting case \cite{benA90} is when apart from
just diffusion, only coagulation and decoagulation are present, that is
$w_{1,1}(0,0)=w_{1,0}(0,0)=0$. We choose
\BEQ
\frac{\vph^{(i)}(1,1)}{\vph^{(i)}(1,0)} =
\sqrt{ \frac{w_{1,0}(1,1)}{w_{1,0}(0,1)} } \;\; , \;\;(i = 1, 2, ..., L-1).
\EEQ
We stress that this transformation is singular when the decoagulation rate
$w_{1,0}(1,1)$ goes to zero. We get a hermitian Hamiltonian
$H= H_0 + H_1$ with  $H_0$ being given by Eq.~(\ref{gl4.21}) with
\BEA
\Delta' &=& 1 - {w_{1,0}(0,1)} \\
\Delta &=& \Delta' + {w_{1,0}(1,1)}
\EEA
and
\BEQ
H_1 = - \sum_{i=1}^{L-1} \sqrt{ (\Delta - \Delta')(1 - \Delta') }
\left( \sig_{i}^{x} E_{i+1}^{11} + E_{i}^{11} \sig_{i+1}^{x} \right)
\EEQ
where $\Delta' < \Delta$ and again we suspect a massive phase, at least
when $\Delta'$ is close to $\Delta$. Note that now $H_1$ is hermitian and its
presence does change the spectrum of $H$. This change is in fact quite
important. If we look directly at the master equation, besides the
trivial stationary probability distribution Eq.~(\ref{gl2.13}) there
is a second one which satisfies $\partial_t P = 0$ and is given by
\BEQ \label{gl4.31}
P(\{\beta\};t) = \prod_{k=1}^{L} p(\beta_k )
\EEQ
where
\BEA
p(\beta) &=& \exp\left( \mu (-1)^{\beta} + \nu \right) \nonumber \\
\mu &=& \ln \sqrt{ \frac{w_{1,0}(0,1)}{w_{1,0}(1,1)} } \;\; , \;\;
\nu = - \mu -\ln \left( 1 + \frac{w_{1,0}(1,1)}{w_{1,0}(0,1)} \right)
\EEA
The probability distribution Eq.~(\ref{gl4.31}) corresponds to
a one-dimensional Ising model defined on the dual lattice and the
site variable $\beta_k$ corresponds to a link variable of the dual lattice,
see also Section~\ref{sect7}.

The existence of two stationary probability distributions
satisfying $\partial_{t} P =0$ makes
the Hamiltonian $H$ have a degenerate vacuum.
The existence of a decoagulation process, no matter how small it is, implies,
if the annihilation process is absent,
that the stationary configuration should
correspond to Eq.~(\ref{gl4.31}), unless the
initial state is the empty lattice.
If we now turn on the annihilation mechanism, the final configuration will
be the empty lattice. Since we always have
massive behaviour this transition between
stationary states will correspond to a first-order nonequilibrium
phase transition which may be accounted by the fact that no symmetry of $H$
is broken.
This can now be studied by performing perturbative
calculations in the limit of a small decoagulation rate.

We are aware of the following fascinating puzzle. Although the Hamiltonian
describing the coagulation and decoagulation processes is not
known to be integrable it was shown in \cite{benA90} that
the gap-probability function for this
model can be computed exactly and this is
an indication that the model is integrable. This is at least the case if one
considers the creation-annihilation model with diffusion which
will be discussed next (recall that
Glauber's solution \cite{Glau63} was discovered without
using the Jordan-Wigner transformation). We will return to this problem in
a future publication (see also Section~\ref{sect5}).

\subsection{$Z_2$ and parity invariant, not vacuum-driven processes}

The full master equation without any restrictions includes besides the
processes considered so far the following two, see Eq.~(\ref{rel4.2})
\begin{enumerate}
\item[g.)] creation, with rate $w_{1,1}(1,1)$
\BEQ
\emptyset + \emptyset \rar A + A
\EEQ
\item[h.)] birth, with rate $w_{1,0}(1,0)$
\BEQ
\emptyset + \emptyset \rar A + \emptyset
\EEQ
\end{enumerate}
{}From probability conservation, we have
\BEA
w_{0,0}(0,0) &=& w_{1,1}(1,1) + 2 w_{1,0}(1,0) \nonumber \\
w_{0,0}(1,0) &=& w_{1,0}(0,0) + w_{1,0}(1,1) + w_{1,1}(1,0) \\
w_{0,0}(1,1) &=& 2 w_{1,0}(0,1) + w_{1,1}(0,0) \nonumber
\EEA
The Hamitonian is
\BEA \label{hha}
H &=& \sum_{i=1}^{L-1} \left[ \frac{1}{4} \left( w_{0,0}(0,0)
+ 2 w_{0,0}(1,0) + w_{0,0}(1,1) \right) \right. \nonumber \\
& & + \frac{1}{4} \left( w_{0,0}(0,0)+w_{0,0}(1,1) -2 w_{0,0}(1,0) \right)
\sig_{i}^{z} \sig_{i+1}^{z} \nonumber \\
& & + \frac{1}{4} \left( w_{0,0}(0,0) - w_{0,0}(1,1) \right)
\left( \sig_{i}^{z} + \sig_{i+1}^{z} \right) \nonumber \\
& & -w_{1,1}(1,0) \left( E_{i}^{01} E_{i+1}^{10} + E_{i}^{10} E_{i+1}^{01}
\right) \nonumber \\
& & - \lambda^2 w_{1,1}(0,0) E_{i}^{01} E_{i+1}^{01}
    - \lambda^{-2} w_{1,1}(1,1) E_{i}^{10} E_{i+1}^{10}
\nonumber \\
& & - \lambda w_{1,0}(0,0)
\left( E_{i}^{01}E_{i+1}^{00} + E_{i}^{00}E_{i+1}^{01} \right)
\nonumber \\
& & - \lambda^{-1} w_{1,0}(1,0)
\left( E_{i}^{10}E_{i+1}^{00} + E_{i}^{00}E_{i+1}^{10} \right)
\nonumber \\
& & - \lambda w_{1,0}(0,1)
\left( E_{i}^{01}E_{i+1}^{11} + E_{i}^{11}E_{i+1}^{01} \right)
\nonumber \\
& & - \left. \lambda^{-1} w_{1,0}(1,1)
\left( E_{i}^{10}E_{i+1}^{11} + E_{i}^{11}E_{i+1}^{10} \right) \right],
\nonumber \\ & &
\EEA
where we have choosen $\frac{h^{(k)}(1)}{h^{(k)}(0)} =
\lambda, (k = 1, 2, ..., L)$
in Eq.~(\ref{alela}).
This Hamiltonian has an interesting property. It can always be brought to a
hermitian form
through a similarity transformation if one out of the
three possible forward-backward reactions (annihilation-creation,
death-birth, coagulation-decoagulation) take place. If, however,
the following relation between the rates is satisfied
\BEQ \label{rel4.47}
\frac{w_{1,1}(1,1)}{w_{1,1}(0,0)} = \left( \frac{w_{1,0}(1,1)}{w_{1,0}(0,1)}
\right) ^2 =
\left( \frac{w_{1,0}(1,0)}{w_{1,0}(0,0)}\right) ^2
\EEQ
$H$ can be made hermitian even when all three pairs of reactions occur.
The physical significance of Eq.~(\ref{rel4.47}) has still to be explored.

We shall confine ourselves to the $Z_2$ preserving processes which
include besides diffusion only annihilation and creation processes.
We make the transformation
\BEQ \label{gl4.38}
\frac{\vph^{(k)}(1,1)}{\vph^{(k)}(0,0)} =
\sqrt{ \frac{w_{1,1}(1,1)}{w_{1,1}(0,0)} } \;\; , \;\; (k = 1, 2, ..., L-1),
\EEQ
and find $H/w_{1,1}(1,0) = H_0 + H_1$ where
\BEA
H_0 &=& -\frac{1}{2} \sum_{i=1}^{L-1} \left[ \sig_{i}^{x} \sig_{i+1}^{x}
+ \sig_{i}^{y} \sig_{i+1}^{y} + \Delta \sig_{i}^{z} \sig_{i+1}^{z}
+ (1 - \Delta') \left( \sig_{i}^{z} + \sig_{i+1}^{z} \right)
+ \Delta -2 \right] \nonumber \\
H_1 &=& -\frac{1}{2} \sum_{i=1}^{L-1} \sqrt{(\Delta' - \Delta)
(2-\Delta-\Delta' )} \left( \sig_{i}^{x} \sig_{i+1}^{x} -
\sig_{i}^{y} \sig_{i+1}^{y} \right) \label{gl4.41}
\EEA
\BEA
\Delta' &=& 1 - \frac{w_{1,1}(0,0) - w_{1,1}(1,1)}{2 w_{1,1}(1,0)} \\
\Delta &=& \Delta' - \frac{w_{1,1}(1,1)}{w_{1,1}(1,0)}
\EEA
This is just the Hamiltonian corresponding to the kinetic Ising
model with purely relaxational dynamics \cite{Ref20}
and $H$ is hermitian. The limit
when the elements of $H_1$ are small does need some care. If the
annihilation and creation rates are equal, $\Delta' =1$ which allows us to do
perturbation theory in the coupling $1-\Delta$ of Eq.~(\ref{gl4.41}).
However, if the rates are not equal and one just takes $w_{1,1}(1,1)$ to zero,
the limit for the eigenvectors  is {\em singular} as is the transformation
Eq.~(\ref{gl4.38}), although the eigenvalue spectrum is not
affected. Therefore, the heuristically appealing relationship \cite{Ref21}
between the kinetic Ising model with Glauber dynamics and the
annihilation process should be considered with care if one works (like in
Ref. \cite{Ref20}) with the
hermitian formulation given by Eq. ~(\ref{gl4.41}).

\section{Exact solution of a two-state system}\label{sectA}

In this section, we shall discuss the dynamics of the two-states system
described by the Hamiltonian Eq.~(\ref{gl4.8}) in the free fermion case
$\Delta=0$. (For a connexion between this  Hamiltonian and
a seven-vertex model, see Appendix B.)
This model describes $A$ particles diffusing and
annihilating in pairs. The motivation for treating this particular
case in great detail is twofold.

Firstly, we would like to show that a finite-size scaling theory
can be formulated for nonequilibrium systems. This point is particularly
important if (in contradistinction to this example) the problem is not
exactly solvable. In that case, one will have to solve the problem
numerically for finite chains and then extrapolate the results
to infinite systems, as one does for equilibrium problems \cite{Priv90}
(see also Appendix A).
One expects that the finite-size scaling proven for this
model is of general validity.

For instance, the concentration of particles per site $<n>/L$ obeys the
relation
\BEQ \label{eq:DFFS}
c =  L^{x} \phi (z)
\EEQ
where $z=4Dt/L^{2}$ and $\phi(z)\sim z^{-\alpha}$ as $z\rar 0$. Moreover,
$x$ is related to the large time behaviour of $c \sim t^{-\alpha}$,
namely $x=-2\alpha$.

Secondly, we want to stress the interest of the (connected) two-point function
\BEQ \label{eq:CCFN}
G(r,t) = <n_m n_{m+r}> - <n_m> <n_{m+r}>
\EEQ
where $m$ indicates the position of a lattice site. Here, we shall pose two
different questions about $G(r,t)$. First, we ask for the
large time behaviour ($t\rar\infty$) if $r$ is kept fixed and look for the
exponent $v_r$
\BEQ \label{eq:VR}
G(r,t) \sim t^{v_{r}}
\EEQ
Secondly, we consider the scaling limit where simultaneously $r,t\rar\infty$
such that $u=r^2/t$ is fixed and define a critical exponent $y$
\BEQ \label{eq:YExp}
G(r,t) \sim r^{-y} g(u)
\EEQ
where $g$ is a scaling function. We believe that for nonequilibrium problems
the exponents $y$'s might
turn out to be as fundamental as the critical exponents of two-dimensional
equilibrium statistical mechanics, since the information given by the
initial probability function should be hidden in the scaling function
$g(u)$ and consequently the $y'$s should be universal (see also Ref.
\cite{Ref21}).

Finally, we would like to present powerful techniques which allow us
to reduce the calculation of nonequilibrium averages, see eq.~(\ref{rel3.18}),
to vacuum expectation values, thereby extending
the approach initiated by Lushnikov \cite{Lush86} to the calculation
of correlation functions.

The quantum Hamiltonian $H$ ~(\ref{gl4.8}) contains the parameter
\BEQ
\eps = \frac{1}{2}\left( q + q^{-1}\right) -1 \geq 0
\EEQ
where $\eps$ measures the left-right asymmetry of the diffusion
process. In the remainder of this section we take $\eps=0$.
We do not make use of the change of variables given by the function
$\vph^{(i)}(\alpha,\beta)$ described in the last section. In terms of
Pauli matrices $\sig^{\pm} = \left( \sig^{x} \pm i \sig^{y} \right) /2$,
the Hamiltonian is equivalent to
\BEQ \label{glA2}
H = - D \sum_{n=1}^{L} \left[ \sig_{n}^{+} \sig_{n+1}^{-} +
\sig_{n}^{-} \sig_{n+1}^{+} - 2 \sig_{n}^{+} \sig_{n}^{-}
+ 2 \sig_{n}^{-} \sig_{n+1}^{-} \right]
\EEQ
where $D$ is the diffusion constant. In this section for simplicity we only
use periodic boundary conditions.

\subsection{Diagonalization and the generating function}

The diagonalization $H$ is fairly standard. By restricting ourselves to
the case $L=2M$ even we  define fermionic variables via
the Jordan-Wigner transformation
\BEA
\sig_{n}^{-} &=& ~ ~ ~ (-1)^{\sum_{m<n} a_{m}^{+} a_{m}} \, a_{n} \nonumber \\
\sig_{n}^{+} &=& a_{n}^{+} \, (-1)^{\sum_{m<n} a_{m}^{+} a_{m}} \label{glA3}
\EEA
as well as their Fourier transforms
\BEA
a_{n} &=& L^{-1/2} e^{-i \pi /4} \sum_{q} e^{i qn} \, \tilde{a}_q \nonumber \\
a_{n}^{+} &=& L^{-1/2} e^{+i \pi /4} \sum_{q} e^{-i qn} \, \tilde{a}_{q}^{+}
\label{Four}
\EEA
In terms of these fermionic variables the Hamiltonian Eq.~(\ref{glA2})
takes the form
\BEA
H &=& -2D \sum_{0\leq q < \pi}
\left[ \left( \cos q -1 \right) \left( \tia{q}^{+}
\tia{q} + \tia{-q}^{+}\tia{-q} \right) + 2 \sin q \, \tia{q}\tia{-q}
\right. \nonumber \\
&& - \left. \delta_{q,\pi} 2
\tia{\pi}^{+}\tia{\pi} \right] \label{glA5}
\EEA
where $q$ takes the values
\BEA
q_{\mbox{\rm even}} &=& \frac{2k-1}{2M}\pi \;\; , \;\; k=1,2,\ldots,M
\nonumber \\
q_{\mbox{\rm odd}} &=& \frac{2k}{2M}\pi \;\; ,
\;\; k=0,1,\ldots, M \label{glA6}
\EEA
if the total number of fermions ${\cal N} = \sum_{m=1}^{L}
a_{m}^{+} a_{m}$ is an even or odd number, respectively. The Hilbert space
associated to Eq.~(\ref{glA5}) can therefore be splitted into two
disjoints sectors depending on the parity of the   fermion number. The sector
with ${\cal N}$ odd (${\cal N}$ even) will be related to the case where
the ground state has a single particle (no particle).

Looking again at Eq.~(\ref{glA5}), we note that $H = \sum_{0 \leq q \leq \pi}
 H_q$
has block-diagonal structure. The time-evolution equation for wave functions
$\partial_{t} \Phi = -H \Phi$ can be broken up into separate equations
for each block
\BEQ \label{glA7}
\Phi(t) = \prod_{0 \leq q \leq \pi} \Phi_{q}(t) \;\; , \;\;
\partial_t \Phi_q = -H_q \Phi_q; (0 \leq q \leq \pi).
\EEQ
Each block is generated by acting with $\tia{q}^{+}$ and $\tia{-q}^{+}$
on the state $\ket{vac}$ without fermions, i.e. $\tilde{a}_{q}\ket{vac}=0$.
If $q \neq 0,\pi$, $H_q$ can be written as a $4 \times 4$ matrix. In the basis
$\ket{0} = \ket{vac}, \ket{1} = \tia{q}^+ \ket{vac}, \ket{2} = \tia{-q}^+
\ket{vac}$
and $\ket{3} = \tia{q}^+ \tia{-q}^+ \ket{vac}$ it takes the form
\BEQ \label{glA8}
H_q = -2D \left( \begin{array}{cccc}
0 & & & -2 \sin q \\
 & \cos q \, -1  & & \\
 & & \cos q \, -1 & \\
 & & & 2\cos q \, -2 \end{array} \right)
\EEQ
while for $q=0,\pi$, the blocks $H_0$ and $H_{\pi}$ are already diagonal.
The general solution of Eq.~(\ref{glA7}) corresponds to
\BEQ \label{glA}
\Phi_{q}(t) = \left( \alpha_{q}(t) \tia{q}^{+}\tia{-q}^{+}
+\gamma_{q}(t)\tia{q}^{+} + \delta_{q}(t) \tia{-q}^{+} +\beta_{q}(t)
\right) \ket{vac}
\EEQ
where the functions $\alpha,\beta,\gamma,\delta$ satisfy the
differential equations
\BEA
\dot{\alpha}_{q}(t) &=& 4D \left( \cos q \, -1 \right) \alpha_{q}(t)
\nonumber \\
\dot{\beta}_{q}(t) &=& -4D\sin q \, \alpha_{q}(t) \nonumber \\
\dot{\gamma}_{q}(t) &=& 2D \left( \cos q \, -1 \right) \gamma_{q}(t)
\nonumber \\
\dot{\delta}_{q}(t) &=& 2D \left( \cos q \, -1 \right) \delta_{q}(t)
\EEA
The solutions of these equations are promptly derived
\BEA
\alpha_{q}(t) &=& \alpha_{q}(0) \exp \left[ -4Dt \left( 1 -\cos q
\right) \right] \nonumber \\
\beta_{q}(t) &=& \beta_{q}(0) + \cot \frac{q}{2} \,
\left\{ \alpha_{q}(t) -^{ } \alpha_{q}(0) \right\} \nonumber \\
\gamma_{q}(t) &=& \gamma_{q}(0) \exp \left[ -2Dt \left( 1 -\cos q
\right) \right] \nonumber \\
\delta_{q}(t) &=& \delta_{q}(0) \exp \left[ -2Dt \left( 1 -\cos q
\right) \right]
\EEA
For an arbitrary initial condition the general solution will be given
by linear combinations of the wave functions Eq.~(\ref{glA7}). A general
study considering arbitrary
general initial conditions is not straightforward and is
not in our intentions. Since our main interest here
is the finite-size scaling behaviour of the system, we will hereafter
choose the simple initial condition where we have no vacancies
present. This condition was used by Lushnikov \cite{Lush86} and corresponds
to the situations where we have a single wave function (\ref{glA7}),
i.e. $\psi(t)=\Phi(t)$ with
\BEQ \label{AC16}
\alpha_{q}(0)=-1 \;\; , \;\; \beta_{q}(0) = \gamma_{q}(0)=\delta_{q}(0)=0,
\EEQ
where $q$ takes the values $q_{even}$ in the set (\ref{glA6}), since in this
case the fermion number ${\cal N}$ is even.

In order to calculate the concentration of particles and correlations,
following the work of Lushnikov \cite{Lush86} it is convenient to
introduce  generating functions. Since (see Section~\ref{sect3})
the wave function takes the role of the probability distribution
introduced earlier, a generating function convenient for our
purpose is defined by
\BEQ \label{glA12}
F(\{z\},t) = F(z_1,\ldots,z_L,t) = \frac{ \sum_{\{{\cal C}\}}
\prod_{i=1}^{L} z_{i}^{n_i}
\psi(\{{\cal C}\};t) }{ \sum_{\{{\cal C}\}} \psi(\{{\cal C}\};t) }
\EEQ
where $n_i =0,1$ and $z_i$ denote the occupation number and fugacity
at the site $i$, ${\cal C}$ is a configuration
$\{n_1,\ldots,n_L\}$ of occupied or
empty sites. In Eq.~(\ref{glA12}) we take into account
that $\psi$ does not, in general, satisfy the normalization
condition of a probability distribution. Using the results of
Section~\ref{sect3}, in particular (3.14), we can bring the
calculation of $F(\{z\},t)$
to an equilibrium problem, i.e.
\BEA
F(\{z\},t) &=& F_{0}(t) \sum_{\{ {\cal C}\} } \prod_{i=1}^{L} z_{i}^{n_i}
\psi( \{ {\cal C} \};t) \label{eq:GenFun} \\
&=& F_{0}(t) \bra{vac} \exp\left(
\sum_{\ell=1}^{L} z_{\ell} \sig^{-}(\ell)\right)
\prod_{q>0} \left( \alpha_{q}(t) \tia{q}^{+}\tia{-q}^{+} + \beta_{q}(t)
\right) \ket{vac} \label{eq:vacvac}
\EEA
where we have already used the initial condition Eq.~(\ref{AC16})
and $F_{0}(t)$ is determined
from the normalization condition $F(1,\ldots,1,t)=1$ for all $t$.
In order to understand Eq.~(\ref{eq:vacvac}), it is useful to consider the
identity
\BEQ
\exp \left( \sum_{m=1}^{L} z_m \sig^{-}(m) \right) =
\sum_{k=0}^{L}  \sum_{m_1 > m_2 > \ldots > m_k} z_{m_{1}}
\sig^{-}(m_1) \cdots z_{m_{k}} \sig^{-}(m_k)
\EEQ
which we can derive by expanding the exponentials and using
the relations $\left( \sig^{-}(m)\right)^{2} =0$ and
$\left[ \sig^{-}(m), \sig^{-}(m')\right] =0$. It is now clear that
$\bra{vac}\exp\left( \sum_{\ell=1}^{L} z_{\ell} \sig^{-}(\ell) \right)$
corresponds to the state $\bra{0}$ of Section~\ref{sect3}.

In order to calculate $F(\{z\},t)$ we observe that due to the ordering
$m_1 > m_2 > \ldots > m_k$ we get
\BEQ \label{glA14}
\bra{vac} \sig^{-}(m_1) \sig^{-}(m_2) \dots \sig^{-}(m_k)  = \bra{vac}
a_{m_1} a_{m_2} \ldots a_{m_k}
\EEQ
On the other hand, we also have from Eq.~(\ref{Four})
\BEQ
\tia{q}^{+} \tia{-q}^{+} = \frac{2}{L} \sum_{m} \sum_{n > m} \sin (q(n-m)) \,
a_{n}^{+} a_{m}^{+}
\EEQ
Because of the product structure Eqs.~(\ref{glA7}, \ref{glA}) of
$\phi$, the generating function can also be written as
$F = \prod_{q>0} F_q$. Each factor contains only two fermionic creation
operators at most. So we only need to calculate
\BEA
\lefteqn{ \sum_{m'}\sum_{m>m'} \bra{vac} \sig^{-}(m) \sig^{-}(m') \tia{q}^{+}
\tia{-q}^{+} \ket{vac} } \nonumber \\
&=& \frac{2}{L} \sum_{n',m'} \sum_{m>m'} \sum_{n>n'} \left[ \sin (q(n-n')) \,
\bra{vac} a_m a_{m'} a_{n}^{+} a_{n'}^{+} \ket{vac} \right] \nonumber \\
&=& \frac{2}{L} \sum_{n',m'} \sum_{m>m'} \sum_{n>n'} \left[ \sin (q(n-n')) \,
\left( \delta_{n,m'} \delta_{n',m} - \delta_{n,m} \delta_{n',m'} \right)
\right]  \nonumber \\
&=& - \frac{2}{L} \sum_{n'}\sum_{n>n'} \sin (q(n-n')) \label{glA16}
\EEA
In the sequel, we shall need two identities, which are obtained using
Eq.~(\ref{glA6}), with $m$ fixed
\BEA
\sum_{n>m} \sin (q(n-m)) &=& \frac{1}{2} \left[ \cot \frac{q}{2} +
\frac{\cos (q(m-1/2))}{\sin (q/2)} \right] \nonumber \\
\sum_{n<m} \sin (q(m-n)) &=& \frac{1}{2} \left[ \cot \frac{q}{2} -
\frac{\cos (q(m-1/2))}{\sin (q/2)} \right] \label{soseno}
\EEA

We now combine Eqs.~(\ref{eq:GenFun}) - (\ref{soseno}) and get
\BEQ
F(\{z\},t) = F_{0}(t) \prod_{q>0} \left(  \alpha_{q}(t) \cdot
\left( -\frac{2}{L}\sum_{m} \sum_{n>m} z_n z_m
\sin (q(n-m)) \right) + \beta_{q}(t) \right)
\EEQ
We fix $F_{0}(t)$ from the condition $F(1,\ldots,1,t)=1$ for all $t$. The final
result for the generating function is
\BEQ \label{eq:FGenFunc}
F(\{z\},t) = \prod_{q>0} \left(  \frac{ \alpha_{q}(t) \cdot
\left( -\frac{2}{L}\sum_{m} \sum_{n>m} z_n z_m
\sin (q(n-m)) \right) + \beta_{q}(t) }{ -\alpha_{q}(t) \cot\frac{q}{2}
+\beta_{q}(t) } \right)
\EEQ

\subsection{Calculation of the mean concentration}

The mean number of particles at an arbitrary site $m$ for a chain of length $L$
is now obtained from
\BEQ \label{AC25}
<n_m> = \left. z_m
\frac{\partial}{\partial z_m}F(\{z\},t) \right|_{z_1=\ldots=z_L=1}
= L^{-1} \sum_{q>0} \left( \frac{-2 \alpha_{q}(t)\cot\frac{q}{2}}{-
 \alpha_{q}(t)\cot\frac{q}{2} + \beta_{q}(t)} \right)
\EEQ
Due to the translational invariance of the particular initial state
considered here (Eq.~(\ref{AC16})) the concentration of particles per site is
simply
\BEQ
c = <n_m>
\EEQ
Substituting Eqs.~(\ref{glA}, \ref{AC16}) into (\ref{AC25}) we obtain
\BEQ
\label{glA21}
c(t) = 2 L^{-1} \sum_{q>0} \exp\left( -4Dt \left(1-\cos q \right) \right)
\EEQ

\subsection{Large-time behaviour and finite-size scaling}

We now discuss the late stages dependence of the concentration $c(t)$.
In the limit $L\rar\infty$ for $t$ fixed
\BEQ
c(t) = \frac{1}{\pi} \int_{0}^{\pi} e^{-4Dt(1-\cos q)} dq
= e^{-4Dt} I_{0}(4Dt)
\EEQ
where $I_{0}(x)$ is a modified Bessel function. For $Dt\gg 1$ one gets
\BEQ
c(t) \simeq (8\pi Dt)^{-1/2}
\EEQ

Let us now analyse the regime of finite-size scaling where
$L\rar\infty$, $t\rar\infty$ but $tL^{-2}$ fixed. In this case we obtain
\BEA
c &=& \frac{2}{L}
\sum_{k=1}^{\infty} \exp \left( -2\pi^{2} z \left( k -\frac{1}{2}
\right)^{2} \right) \label{AC30a} \\
&=& \frac{1}{L} \Theta_{2}(0,2\pi i z) \label{AC30b} \\
&=& \frac{1}{L} \sqrt{\frac{1}{2\pi z}} \left( 1+ 2\sum_{\ell=1}^{\infty}
(-1)^{\ell} \exp\left( -\ell^{2}/(2z) \right) \right) \label{AC30c}
\EEA
where $\Theta_{2}(z,\tau)$ is a Jacobi theta function and
\BEQ \label{glA26}
z = 4 D t L^{-2}
\EEQ
is the finite-size scaling variable. Eq.~(\ref{AC30c}) is obtained from
(\ref{AC30a}) by using the Poisson resummation formula.
{}From Eqs.~(\ref{AC30a}-\ref{AC30c}) and (\ref{eq:DFFS}) we get the critical
exponent $x=-1$.

\subsection{Fluctuations around the mean values}

Next, we briefly discuss the fluctuations around $<n_m>$. For simplicity,
we take $z_1=\ldots=z_L=z$ in Eq.~(\ref{glA12}). Then the generating function
$F=F(z,t)$ and we have
\BEA
<n_{m}^{2}> &=& \left.
\frac{1}{L^2} \left( z \frac{\partial}{\partial z}\right)^{2}
F(z,t)\right|_{z=1} \nonumber \\
&=& <n_m>^{2} + 2<n_m>/L -
\frac{1}{L^2}\sum_{q>0}\left(\frac{ -2\alpha_{q}(t)\cot \frac{q}{2}}{
-\alpha_{q}(t)\cot \frac{q}{2}  + \beta_{q}(t)} \right)^{2}
\EEA
We get
\BEA
<n_{m}^{2}> - <n_m>^{2} &=& 2/L \left( e^{-4Dt} I_{0}(4Dt) - e^{-8Dt}
I_{0}(8Dt)
\right) \nonumber \\
&\simeq& \frac{1}{L}
\frac{1}{\sqrt{2\pi D t}} \left( 1 - \frac{1}{\sqrt{2}} \right)
\EEA
as $t\rar\infty$. This means that for fixed, but large $t$
\BEQ
<\delta n_{m}^{2}>/<n_m>^{2} \sim t^{1/2} L^{-1}
\EEQ
which accounts for the fact that as the time grows the annihilation
of couples of particles induces larger fluctuations in the particle
concentration.

\subsection{Correlation functions}

We now turn towards the calculation of correlation functions. As
an example which illustrates the general technique, we calculate the
connected correlation function defined in Eq.~(\ref{eq:CCFN}). From
the definition of the generating function (\ref{glA12}) we can
write
\BEQ
G(r,t) = <n_{m} n_{m+r}>-<n_m>^{2} = \left.
\frac{\partial^2}{\partial z_m \partial z_{m+r}}
\ln F(\{z\},t)\right|_{z_1=\ldots=z_L =1}
\EEQ
A straightforward calculation gives us
\BEQ \label{glA42}
G(r,t)=\frac{2}{L} \sum_{q>0} \left( \frac{ -\sin (qr) \,
\alpha_{q}(t)}{-\alpha_{q}(t)\cot \frac{q}{2}+\beta_{q}(t)}
\right)
-\frac{4}{L^2}\sum_{q>0}\left(\frac{-\alpha_{q}(t)\cot\frac{q}{2}}{-
\alpha_{q}(t)\cot\frac{q}{2}+\beta_{q}(t)}\right)^{2}
\EEQ
which does not depend on $m$ due to the fact that the Hamiltonian
Eq.~(\ref{glA2}) as well as the initial probability distribution
are translational invariant.

Analysing the relative importance of the two terms contributing to
$G(r,t)$, we see that in the limit $L\rar\infty$, the first term
is of order unity, while the second one is of order ${\cal O}(L^{-1})$.

This correlation can be evaluated in the same way as we did for the
concentration. Let us consider initially the case of the correlation
of next-neighbour particles. We find
\BEA
G(1,t) &=& \frac{2}{L} \sum_{q>0} \left( 1 -\cos q\right)
\exp \left[ -4Dt \left( 1-\cos q\right) \right] \nonumber \\
&=& e^{-4Dt} \left[ I_{0}(4Dt) - I_{1}(4Dt) \right] \nonumber \\
&\simeq& \pi (8\pi Dt)^{-3/2}
\EEA
as $t\rar\infty$. The same result can be obtained using Glauber's
dynamics for the Ising model at $T=0$ (see also Section~\ref{sect7})
and considering domain walls two-point functions \cite{Ref21a}.
The initial condition we are considering in the diffusion-annihilation
problem (no vacancies) corresponds to a fully ordered
antiferromagnetic state in the kinetic Ising model. On the other
hand, for arbitrary but finite $r$, we obtain analogously
\BEQ
G(r,t) \simeq r G(1,t)
\EEQ
This result reflects the fact that at late times it is more
difficult to find particles close to each other than far apart.
{}From Eq.~(\ref{eq:VR}) this gives us the exponents $v_r = -3/2$.

Finally, in connection with Eq.~(\ref{eq:YExp}), let us consider
the limit where $r\rar\infty$, $t\rar\infty$ but
$u= r^{2}/t$ stays finite. This leads to
\BEQ \label{eq:Ende}
G \simeq \pi (8\pi D)^{-3/2} r^{-2} u^{3/2} e^{-u/(2D)}
\EEQ
and we read off the exponent $y=-2$ and recognize the scaling
function $g(u) =\pi (8\pi D)^{-3/2} u^{3/2} e^{-u/(2D)}$,
see Eq.~(\ref{eq:YExp}).
We remark that if we consider the domain
walls two-point function in the zero-temperature kinetic Ising model
with Glauber dynamics \cite{Ref21a} for an initial
probability distribution with zero magnetization (which is a different
initial condition than that we are considering here), we obtain the
same result. This is an indication in favour of our conjecture that the
$y$ exponents are universal.

\section{Three-state models with $Z_3$ symmetry and va\-cu\-um-dri\-ven
pro\-ces\-ses} \label{sect5}

As noticed in Section~\ref{sect4}, behind many processes in the two-states
models there is the XXZ model in a magnetic field. We shall now show
that there are several integrable quantum chains which play a similar role for
three-states models with $Z_3$ symmetry. We shall restrict
ourselves to the $Z_3$-symmetric
case for simplicity. This already contains the two-species
annihilation process but not the $Z_2$ symmetric trapping reaction to
which we hope to return in a future publication. Let $0$ stand
for a vacancy, $1$ for the particle $A$ and $2$ for the particle $B$.

We use the general results of Section~\ref{sect3} and the conditions
Eqs.~(\ref{gl2.6},\ref{gl2.12}) and derive the Hamiltonian
\BEQ
H = U - T
\EEQ
\BEA
T &=& \sum_{i=1}^{L-1} \left[
 w_{2,1}(0,0) E_{i}^{02} E_{i+1}^{01} +
 w_{1,2}(0,0) E_{i}^{01} E_{i+1}^{02}
\right. \nonumber \\
& &+ w_{2,1}(1,0) E_{i}^{10} E_{i+1}^{01} +
 w_{1,2}(0,1) E_{i}^{01} E_{i+1}^{10} \nonumber \\
& &+ w_{2,1}(2,1) E_{i}^{21} E_{i+1}^{12} +
 w_{1,2}(1,2) E_{i}^{12} E_{i+1}^{21} \nonumber \\
& &+ w_{2,1}(0,2) E_{i}^{02} E_{i+1}^{20} +
 w_{1,2}(2,0) E_{i}^{20} E_{i+1}^{02} \nonumber \\
& &+ w_{2,1}(1,1) E_{i}^{10} E_{i+1}^{12} +
w_{1,2}(0,2) E_{i}^{01} E_{i+1}^{21} \nonumber \\
& &+ w_{2,1}(2,0) E_{i}^{21} E_{i+1}^{01} +
w_{1,2}(1,1) E_{i}^{12} E_{i+1}^{10} \nonumber \\
& &+ w_{2,1}(0,1) E_{i}^{02} E_{i+1}^{12} +
w_{1,2}(2,2) E_{i}^{20} E_{i+1}^{21} \nonumber \\
& &+ w_{2,1}(2,2) E_{i}^{21} E_{i+1}^{20} +
w_{1,2}(1,0) E_{i}^{12} E_{i+1}^{02} \nonumber \\
& &+ \left. w_{2,1}(1,2) E_{i}^{10} E_{i+1}^{20} +
w_{1,2}(2,1) E_{i}^{20} E_{i+1}^{10} \right]
\label{gl5.2a} \\
U &=& \sum_{i=1}^{L-1} \left[
\left( w_{2,1}(2,0) + w_{1,2}(0,2) \right) E_{i}^{11} E_{i+1}^{11}
+ \left( w_{2,1}(0,1) + w_{1,2}(1,0) \right) E_{i}^{22} E_{i+1}^{22}
\right. \nonumber \\
& &+ \left( w_{2,1}(1,0) + w_{1,2}(2,2) \right) E_{i}^{00} E_{i+1}^{11}
+ \left( w_{2,1}(2,2) + w_{1,2}(0,1) \right) E_{i}^{11} E_{i+1}^{00}
\nonumber \\
& &+ \left( w_{2,1}(1,1) + w_{1,2}(2,0) \right) E_{i}^{00} E_{i+1}^{22}
+ \left( w_{2,1}(0,2) + w_{1,2}(1,1) \right) E_{i}^{22} E_{i+1}^{00}
\nonumber \\
& &+ \left( w_{2,1}(2,1) + w_{1,2}(0,0) \right) E_{i}^{11} E_{i+1}^{22}
+ \left( w_{2,1}(0,0) + w_{1,2}(1,2) \right) E_{i}^{22} E_{i+1}^{11}
\nonumber \\
& &+ \left. \left( w_{2,1}(1,2) + w_{1,2}(2,1) \right) E_{i}^{00} E_{i+1}^{00}
\right]. \label{gl5.2b}
\EEA
Notice that for the time being we do not use any equivalence transformation
(\ref{eq:2.8}) as we did in Sec.(\ref{sect4}).

We first consider the case of asymmetric diffusion
\BEQ \label{eq:5.5}
\left\{ \begin{array}{cc}
\emptyset + A \rar A + \emptyset & \mbox{\rm with rate $w_{2,1}(1,0)$} \\
A + B \rar B + A & \mbox{\rm with rate $w_{2,1}(2,1)$} \\
B + \emptyset \rar \emptyset + B & \mbox{\rm with rate $w_{2,1}(0,2)$} \\
A + \emptyset \rar \emptyset + A & \mbox{\rm with rate $w_{1,2}(0,1)$} \\
B + A \rar A + B & \mbox{\rm with rate $w_{1,2}(1,2)$} \\
\emptyset + B \rar B + \emptyset & \mbox{\rm with rate $w_{1,2}(2,0)$}
\end{array} \right.
\EEQ
and let us assume that we have
\BEA
w_{2,1}(1,0) &=& w_{1,2}(2,0) = w_{2,1}(2,1) = \Gamma_L \nonumber \\
w_{1,2}(0,1) &=& w_{2,1}(0,2) = w_{1,2}(1,2) = \Gamma_R.
\EEA
By defining
\BEQ
q = \sqrt{ \frac{\Gamma_L}{\Gamma_R} }
\EEQ
we now obtain
\BEA
H &=& D \left\{ \sum_{i=1}^{L-1} \left[ \frac{q+q^{-1}}{2} -
\left( q \sum_{b > a}
 E_{i}^{ab} E_{i+1}^{ba} + q^{-1} \sum_{b < a}
 E_i^{ab}E_{i+1}^{ba} \right. \right. \right. \nonumber \\
& & \left. \left. \left. \frac{q+q^{-1}}{2} \sum_{a=0}^{2} E_{i}^{aa}
E_{i+1}^{aa}
+ \frac{q-q^{-1}}{2}\sum_{a\neq b} \mbox{\rm sign}(a-b)
E_{i}^{aa} E_{i+1}^{bb} \right) \right] \right\} \label{gl5.7a}
\EEA
where $D = \sqrt{\Gamma_L \Gamma_R}$ is the diffusion constant.
Through a similarity transformation $S$, we can bring this Hamiltonian to
the form
\BEA
H' = S H S^{-1} &=& D \left\{ \sum_{i=1}^{L-1} \left[ \frac{q+q^{-1}}{2} -
\left( \sum_{a\neq b}
 E_{i}^{ab} E_{i+1}^{ba} + \frac{q+q^{-1}}{2}\sum_{a=0}^{2} E_{i}^{aa}
E_{i+1}^{aa}  \right. \right. \right. \nonumber \\
& & \left. \left. \left.
+ \frac{q-q^{-1}}{2}\sum_{a\neq b} \mbox{\rm sign}(a-b)
E_{i}^{aa} E_{i+1}^{bb} \right) \right] \right\} \label{gl5.7}
\EEA
This similarity transformation does not have the simple local form of Eq.~
(\ref{eq:2.8}). The argument goes as follows. We first notice that the
Hamiltonian $H'$ given by Eq.(\ref{gl5.7}) is the exactly integrable
$U_qSU(3)$ chain \cite{Ogie87} and corresponds
to the anisotropic version of the spin 1 model introduced
by Sutherland \cite{Suth75}, (see also Appendix~C, where the connexion
with the Hecke algebra is shown). On the other hand $H'$ can be brought by
an equivalent transformation to the four-parameter deformation of the $SU(3)$
symmetric chain \cite{Simil}
\BEA
H''&=& \tilde{S} H' \tilde{S}^{-1} = D \left\{ \sum_{i=1}^{L-1}
\left[ \frac{q+q^{-1}}{2} -
\left( f_{10}E_{i}^{10}E_{i+1}^{01} + f_{10}^{-1}E_{i}^{01}E_{i+1}^{10}
\right. \right. \right. \nonumber\\
& & + f_{20}E_{i}^{20}E_{i+1}^{02} + f_{20}^{-1}E_{i}^{02}E_{i+1}^{20} +
f_{21}E_{i}^{21}E_{i+1}^{12} + f_{21}^{-1}E_{i}^{12}E_{i+1}^{21} \nonumber\\
& & \left. \left. \left. +\frac{q+q^{-1}}{2}\sum_{a=0}^{2} E_{i}^{aa}
E_{i+1}^{aa}  + \frac{q-q^{-1}}{2}\sum_{a\neq b} \mbox{\rm sign}(a-b)
E_{i}^{aa} E_{i+1}^{bb} \right) \right] \right\} \label{gl5.7c}
\EEA
We observe that if in Eq.(\ref{gl5.7c}) we take $f_{21} = f_{31} = f_{32} = q$
we recover Eq.(\ref{gl5.7}). The equivalence between the chain given by Eq.(
\ref{gl5.7a}) and the $U_qSU(3)$ chain (\ref{gl5.7}) is similar
to the two-state model in the case
of an asymmetric definition, where we got $U_{q} SU(2)$. The chain
Eq.~(\ref{gl5.7}) is massive unless $q=1$, which corresponds to the
left-right symmetric case and the physical interpretation for these
results is similar to the two species case of Section~\ref{sect4}.
The case where the process $A+B\rar B+A$ is suppressed is, as shown
in Appendix~C, again related to a quotient of a Hecke algebra and
the eigenvalues of the Hamiltonian are the same (the degeneracies may be
different) as for the $U_{q}SU(2)$ symmetric one
given by Eq.~(\ref{eq:purediff}).

We shall take from now on $q=1$ and parity conservation.
We assume that the diffusion rates
are
\BEA
w_{2,1}(1,0) &=& w_{1,2}(0,1) = w_{2,1}(0,2) = w_{1,2}(2,0) = D \nonumber \\
w_{2,1}(2,1) &=& w_{1,2}(1,2) = \lmb D \label{gl5.8}
\EEA
We consider now the annihilation processes
\BEQ \label{rel5.10}
\left\{ \begin{array}{cc}
A + B \rar \emptyset + \emptyset & \mbox{\rm with rate $w_{1,2}(0,0) =
w_{2,1}(0,0) = \alpha_{AB} D$} \\
A + A \rar \emptyset + B & \mbox{\rm with rate $w_{1,2}(0,2) =
w_{2,1}(2,0) = \alpha_{AA} D$} \\
B + B \rar \emptyset + A & \mbox{\rm with rate $w_{1,2}(1,0) =
w_{2,1}(0,1) = \alpha_{BB} D$} \end{array} \right.
\EEQ
In this case, the Hamiltonian has a special form
\BEQ \label{gl5.9}
H = D ( H_0 + H_1)
\EEQ
where
\BEA
H_0 &=& \sum_{i=1}^{L-1} \left[ 2 \alpha_{AA} E_{i}^{11} E_{i+1}^{11}
+ 2\alpha_{BB} E_{i}^{22} E_{i+1}^{22} \right. \nonumber \\
& & + (\lmb + \alpha_{AB}) \left( E_{i}^{11} E_{i+1}^{22} +
E_{i}^{22} E_{i+1}^{11}\right) \nonumber \\
& & + E_{i}^{00} \left( E_{i+1}^{11} + E_{i+1}^{22} \right)
+ \left( E_{i}^{11} + E_{i}^{22} \right) E_{i+1}^{00} \nonumber \\
& & - E_{i}^{10} E_{i+1}^{01} - E_{i}^{01} E_{i+1}^{10}
    - E_{i}^{20} E_{i+1}^{02} - E_{i}^{02} E_{i+1}^{20} \nonumber \\
& & \left. - \lmb \left( E_{i}^{21} E_{i+1}^{12} + E_{i}^{12} E_{i+1}^{21}
\right) \right] \label{gl5.10} \\
\lefteqn{ H_1 = - \sum_{i=1}^{L-1}
\left[ \alpha_{AB} \left( E_{i}^{01} E_{i+1}^{02}
+ E_{i}^{02} E_{i+1}^{01} \right) \right.} \nonumber \\
& & \left. + \alpha_{AA} \left( E_{i}^{21} E_{i+1}^{01}
+ E_{i}^{01} E_{i+1}^{21} \right) +\alpha_{BB}\left( E_{i}^{02}E_{i+1}^{12}
+ E_{i}^{12} E_{i+1}^{02} \right) \right] \label{gl5.11}
\EEA
Using the arguments from Section~\ref{sect4}, one can check that the
spectrum of the hermitian operator $H_0$ coincides with the
one of $H/D$. This Hamiltonian
describes the pure reaction $A + B \rar \emptyset$, if we take
\BEQ \label{gl5.12}
\lmb = \alpha_{AA} =\alpha_{BB} = 0
\EEQ

Let us find cases where the Hamiltonian is integrable. First we look
at the $U_q SU(3)$ chain Eq.~(\ref{gl5.7}) with $q=-1$. If we add to the
Hamitonian fields along the generators of the algebra, the system stays
integrable since the supplementary terms commute with the
integrable $U_q SU(3)$ Hamiltonian. We found only one case where this is
possible. Taking
\BEQ
\lmb =1 \;\; , \;\; \alpha_{AA} = \alpha_{AB} = \alpha_{BB} = 2
\EEQ
we find
\BEQ \label{gl5.14}
H_0 = \sum_{i=1}^{L-1} \left[ -1 + \sum_{a \neq b} E_{i}^{ab} E_{i+1}^{ba}
- \sum_{a=0}^{2} E_{i}^{aa} E_{i+1}^{aa} +
2 \left( \sig_{i}^{0}+ \sig_{i+1}^{0}
\right) \right]
\EEQ
The first three terms in Eq.~(\ref{gl5.14}) give a $SU(3)$ symmetric
Hamiltonian, corresponding to $q=-1$ in Eq.~(\ref{gl5.7a}). The last term
with
\BEQ
\sig^{0} = \left( \begin{array}{ccc}
0 & 0 & 0 \\ 0 & 1 & 0 \\ 0 & 0 & 1 \end{array} \right)
\EEQ
is an external field. This observation should allow us to study the critical
properties of this chemical process.

Another possibility occurs when $\lmb=0$. We can rewrite $H_0$ in the
following form, assuming $\alpha_{AA} = \alpha_{BB}$
\BEA
H_0 &=& \sum_{i=1}^{L-1} \left[ \left( \frac{2\alpha_{AA}+\alpha_{AB}}{2}
-2\right) \sig_{i}^{0}\sig_{i+1}^{0} + \left(
\frac{2\alpha_{AA}-\alpha_{AB}}{2} \right) \sig_{i}^{z}\sig_{i+1}^{z}\right.
\nonumber \\
& &\left. \vekz{~}{ }+\sig_{i}^{0} +\sig_{i+1}^{0} - \tau_{i}^{+}\tau_{i+1}^{-}
-\tau_{i}^{-}\tau_{i+1}^{+} -\rho_{i}^{+}\rho_{i+1}^{-}
-\rho_{i}^{-}\rho_{i+1}^{+} \right] \label{gl5.16}
\EEA
where we have used the notations
\BEQ
\tau^{+} = E^{01} \;\; , \;\; \tau^{-} = E^{10} \;\; , \;\;
\rho^{+} = E^{02} \;\; , \;\; \rho^{-} = E^{20} \;\; , \;\;
\sig^{z} = E^{11} - E^{22} \label{notat}
\EEQ
This Hamiltonian commutes with $\Sigma^{z}$ and $\Sigma^{0}$
where
\BEQ \label{gl5.17}
\Sigma^{z} = \sum_{i=1}^{L} \sig_{i}^{z} \;\; , \;\;
\Sigma^{0} = \sum_{i=1}^{L} \sig_{i}^{0}
\EEQ
For the special choice
\BEQ \label{spec}
\alpha_{AB} = 2 \alpha_{AA}
\EEQ
the Hamiltonian has the larger symmetry $SU(2) \otimes U(1)$ since
$H_0$ commutes also with $\Sigma^{\pm}$ which are defined by
\BEQ
\Sigma^{+} = \sum_{i=1}^{L} \sig_{i}^{+} \;\; , \;\;
\Sigma^{-} = \sum_{i=1}^{L} \sig_{i}^{-}
\EEQ
using $\sig^{+} = E^{12}$ and $\sig^{-} = E^{21}$.

If we finally take $\alpha_{AA}=1$, we have
\BEQ \label{gl5.20}
{H_0} = \sum_{i=1}^{L-1} \left[ \left( \sig_{i}^{0}
+ \sig_{i+1}^{0} \right) - \tau_{i}^{+}\tau_{i+1}^{-}
-\tau_{i}^{-}\tau_{i+1}^{+} -\rho_{i}^{+}\rho_{i+1}^{-}
-\rho_{i}^{-}\rho_{i+1}^{+} \right]
\EEQ
The chain Eq.~(\ref{gl5.20}) is integrable and massless as shown in
Appendix~C. It can also be related to $t-J$ model \cite{tJ} with $J = 0$.
Moreover the spectrum of $H_0$ (not the degeneracies) is that of
a free fermion system. In Appendix C it is also shown that not
only $H_0$ but also $H$ can be expressed in terms of a Hecke algebra
which implies that the system is integrable. This remains valid
even if we don't assume left-right symmetry. The quantum chain given by
Eq.(\ref{gl5.20}) was discovered independently in Refs.\cite{AKL} and
\cite{Gome92}. In the first reference this chain was found as a special case
in a search of chains related to Hecke algebras. In the second reference it
corresponds to a special case of integrable chains defined on nilpotent
representations of the $U_qSU(2)$ quantum group with
$q = \exp {\frac{2 \pi i}{3}}$. In the last reference the $SU(2) \otimes U(1)$
symmetry was observed. This triggered a further investigation by two of the
authors to find the whole symmetry of the chain and obtain its eigenspectrum
\cite{Arnoud}. We would like to
mention that for this system an exact result concerning the
time-dependence of the total concentration $<c_A + c_B>$ is
already known \cite{Priv92}.
Moreover, if we keep the relation (\ref{spec}) but let the parameter
$\alpha_{AA}$ free, one can show that the spectrum (not the degeneracies)
is the same as that of the XXZ model in a field, see Eq.~(\ref{ariro}) with
$\Delta = 1 - \alpha_{AA}$ and $h = 1 - \Delta$, which corresponds to the
Prokrovski-Talapov line of Fig.~1.
Up to now we have found several choices for
the constants $\alpha_{AB}$ and $\alpha_{AA}$ for which $H_0$ is
integrable. We failed to find others, although two other classes of
integrable models, namely the chiral Potts \cite{Ahn} and the one
with $U_{q}SU(2)$ , $q^3=1$ symmetry where
periodic or semiperiodic representations \cite{Dari} are taken, have the
required $Z_3$ symmetry.

Much work is still
to be done on the phase structure of the three-state models. One of
the questions to be asked is the connection between the case
Eq.~(\ref{gl5.12}), where $\alpha_{AA}=0$, which is known to be
massless, to the case $\alpha_{AA}=1$, which is massless as well.

Finally, we mention that in Appendix~C (see Eq.~(\ref{eq:C16})) we give
an example of a $Z_2$ symmetric three-state chain (only
$A+A\rar\emptyset$ reaction) which is also related to a Hecke algebra.

\section{Conditions on the reaction rates for the existence of
steady states} \label{sect6}

Steady states are time-independent solutions of the master equation. In
Section~\ref{sect2} we have seen
that if the reaction rates satisfy the condition
Eq.~(\ref{gl2.12}), then the empty lattice Eq.~(\ref{gl2.13}) is a
steady state. We now ask for the conditions on the rates such
that the probability distribution
\BEQ \label{gl6.1}
{\cal P}(\{\beta\}) = \prod_{k=1}^{L} f(\beta_k)
\EEQ
is a steady state, where $f(\beta)$ is a given arbitrary
function. Our motivation is not only to understand
better the structure of the ground states of the master equation
and of the corresponding quantum Hamiltonians, but also to be able
to address the question of the nonequilibrium behaviour of classical
one-dimensional systems. This amounts to introduce a detailed balance
condition for a prescribed equilibrium condition ${\cal P}$, while in the
previous sections we have always taken the empty lattice as the equilibrium
configuration. Let us explain this point.

Suppose that at temperature $T$ we have a one-dimensional classical
system defined by the probability function
\BEQ \label{gl6.2}
{\cal P}(\{\beta\}) = \prod_{k=1}^{L} g(\alpha_{k} - \alpha_{k+1})
\EEQ
If the $\beta$'s are site variables, the $\alpha$'s are link variables.

In our study we are going to consider the two-state model (Ising model)
where
\BEQ \label{gl6.3}
g_{2}(\beta_{k}) = g_{2}(\alpha_{k}-\alpha_{k+1}) =
\exp \left( (-1)^{\beta_{k}} / T \right)
\EEQ
and the three-state model (chiral Potts model \cite{Ostl81}) where
\BEQ \label{gl6.4}
g_{3}(\beta_{k}) = g_{3}(\alpha_{k}-\alpha_{k+1}) =
\exp \left( \cos\left(\frac{2\pi}{3}\beta_{k}
+ \phi\right)/T \right)
\EEQ
and $\phi$ is a fixed phase. We note the identities
\BEQ
g_{2}(0) g_{2}(1) = 1 \;\; , \;\; g_{3}(0)g_{3}(1)g_{3}(2) =1
\EEQ
If the steady state is given by Eq.~(\ref{gl6.2}), the master equation
describes the evolution of the
one-dimensional classical chain from $P(\{\beta\};t=0) = P_{0}(\{\beta\})$
to the equilibrium probability distribution $P(\{\beta\};t=\infty)
= {\cal P}(\{\beta\})$.

In order to find the conditions on the states, we first make a similarity
transformation Eq.~(\ref{gl2.8}), taking
\BEA
h^{(k)}(\alpha) &=& f^{1/2}(\alpha) \nonumber \\
W_{\ell,m}(\alpha,\beta) &=& \sqrt{ \frac{f(\alpha+\ell)f(\beta+m)}{f(
\alpha) f(\beta)} } w_{\ell,m}(\alpha,\beta)
\EEA
Then the condition to have Eq~(\ref{gl6.1}) reads
\BEQ
w_{0,0}(\alpha,\beta) = {\mathop{{\sum}'}_{\ell,m=0}^{N-1}}
w_{\ell,m}(\alpha,\beta)
\frac{f(\alpha+\ell)f(\beta+m)}{f(\alpha) f(\beta)}
\EEQ
where the prime indicates that the case $\ell=m=0$ should be excluded.
Taking into account the probability conservation condition Eq.~(\ref{gl2.5})
and making the change of functions
\BEQ \label{gl6.7}
w_{\ell,m}(\alpha,\beta) = \frac{\gamma_{\ell,m}(\alpha,\beta)}{f(\alpha+\ell)
f(\beta+m)}
\EEQ
we get the system of equations
\BEQ \label{gl6.8}
{\mathop{{\sum}'}_{\ell,m=0}^{N-1}}
\left( \gamma_{\ell,m}(\alpha,\beta) - \gamma_{\ell,m}
(\alpha-\ell,\beta-m) \right) = 0
\EEQ
We note that in Eq.~(\ref{gl6.8}) there is no longer any explicit reference
to the function $f$. The whole $f$-dependence is in Eq.~(\ref{gl6.7}).

We now write down the independent conditions. To do so, it is
convenient to introduce the following symmetric and antisymmetric
combinations
\BEA
\gamma_{\ell,m}^{s}(\alpha,\beta) &=& \gamma_{\ell,m}(\alpha,\beta)
+ \gamma_{m,\ell}(\beta,\alpha) \nonumber \\
\gamma_{\ell,m}^{a}(\alpha,\beta) &=& \gamma_{\ell,m}(\alpha,\beta)
- \gamma_{m,\ell}(\beta,\alpha)
\EEA
Now, for the two-state system the independent conditions coming from
Eq.~(\ref{gl6.8}) are
\BEQ \label{gl6.9}
\gamma_{1,0}^{s}(0,0) - \gamma_{1,0}^{s}(1,0) =
\gamma_{1,0}^{s}(0,1) - \gamma_{1,0}^{s}(1,1) =
\gamma_{1,1}(1,1) - \gamma_{1,1}(0,0)
\EEQ
\BEQ \label{gl6.10}
\gamma_{1,0}^{a}(1,0) + \gamma_{1,0}^{a}(1,1) + \gamma_{0,1}^{a}(1,0)
+ \gamma_{0,1}^{a}(0,0) = 2 \gamma_{1,1}^{a}(0,1)
\EEQ
Note that Eq.~(\ref{gl6.9}), which relates only symmetric combinations,
does not, as expected, contain the diffusion constant since it fixes the
time scale only. The antisymmetric diffusion combinations appear however
in Eq.~(\ref{gl6.10}) and thus play a dynamical role. We remind the
reader that this property was already noticed in earlier sections
when discussing Hamiltonians with quantum group symmetries ($q\neq 1$).

For the three-state system with $Z_3$ symmetry (which means that only
the functions $w_{1,2}$ and $w_{2,1}$ appear) we have from
Eq.~(\ref{gl6.8}) the following conditions
\BEQ \label{gl6.12}
\gamma_{2,1}^{s}(0,0) = \gamma_{2,1}^{s}(1,2) \;\; , \;\;
\gamma_{2,1}^{s}(1,1) = \gamma_{2,1}^{s}(2,0) \;\; , \;\;
\gamma_{2,1}^{s}(2,2) = \gamma_{2,1}^{s}(0,1)
\EEQ
\BEA
2\gamma_{2,1}^{a}(1,0) &=& \gamma_{2,1}^{a}(2,2) + \gamma_{2,1}^{a}(0,1)
\nonumber \\
2\gamma_{2,1}^{a}(0,2) &=& \gamma_{2,1}^{a}(1,1) + \gamma_{2,1}^{a}(2,0)
\nonumber \\
2\gamma_{2,1}^{a}(2,1) &=& \gamma_{2,1}^{a}(0,0) + \gamma_{2,1}^{a}(1,2)
\label{gl6.13}
\EEA
We observe that the detailed balance equations do not determine all the
rates. The remaining freedom can be used to fit the experimental data.
We shall apply these conditions on the rates in the next two sections.
Before doing so, we briefly recall the questions of interest
and some definitions relevant to
critical dynamics. As already mentioned in the introduction, a classical
Ising spin system has no intrinsic dynamics, because the
Poisson brackets between the spin variables and the Hamiltonian
vanish. In a real magnetic system, the spins are interacting with
other degrees of freedom (phonons, impurities, \ldots) and those interactions
are responsible for the dynamics of the spins. It is very difficult to
implement fully the microscopic dynamics. Accordingly, one replaces
the true dynamics by a fictious one given by a master equation for the
probability that one spin configuration of the system is realized at a
given time. The transition probabilities appearing in the master equation
can in principle be computed from the full microscopic dynamics
(via projective techniques). Unfortunately, it is often not possible
to perform this program, even for simple models \cite{forgas,lage}. Thus, the
transition probabilities are chosen phenomenologically according
to two criteria: i) to satisfy detailed balance, necessary to ensure
that the desired equilibrium state is stationary, ii) to be
qualitatively in agreement with the physics of the system. Once the
dynamics is defined, the main problem is to find how the physical
quantities will relax towards equilibrium. A quantity of particular
interest is the order parameter (the magnetization for the Ising model).
Typically, the order parameter relaxes very slowly in the vicinity of
a second-order phase transition (critical slowing-down). The dynamical
scaling hypothesis assumes that the characteristic time $\tau$ diverges
as a power law of the spatial correlation length $\xi$; $\tau \sim
\xi^{z}$, where $z$ is called the dynamical exponent.

The quantum chain formalism developed here aims to answer two types of
questions:
\begin{enumerate}
\item Does the dynamical scaling hypothesis $\tau \sim\xi^{z}$
always hold ?
\item What is the status of the ``universality'' for the
dynamical exponent $z$ ?
\end{enumerate}

For one-dimensional systems, the critical point is at zero
temperature, $T_c =0$. For small $T$, the inverse spatial
correlation lengths are
\BEA
\xi^{-1} &=& 2 e^{-2/T} \;\; , \;\; ~~~\mbox{\rm two-state system} \nonumber \\
\xi^{-1} &=& \frac{3}{2} \left( \frac{g_{3}(1)}{g_{3}(0)} +
\frac{g_{3}(2)}{g_{3}(0)} \right) \;\; , \;\; ~\mbox{\rm three-state
system}
\EEA
For the three-state system we shall only consider the ordinary Potts model,
where $\phi=0$ and the so-called superintegrable model \cite {Gerit}, where
$\phi=\pi/6$. This serves to illustrate how the dynamics depends on the
equilibrium system. In the sequel, the deviation from the critical
point will be parametrized in terms of the following masses
$\mu \sim \xi^{-1}$, which are related to the temperature as follows
\BEQ
\left\{ \begin{array}{ll}
\mu = e^{-2/T} & \mbox{\rm Ising model} \\
\mu = e^{-3/(2T)} & \mbox{\rm Potts model, $\phi=0$} \\
\mu = e^{-\sqrt{3}/(2T)} & \mbox{\rm superintegrable model, $\phi=\pi/6$}
\end{array} \right.
\EEQ
for the three equilibrium models under consideration. In the next two
sections we shall concentrate on the behaviour for small values of
$\mu$ in the master equation and in the corresponding Hamiltonians.

\section{Critical dynamics for the Ising model} \label{sect7}

We will write the Hamiltonian corresponding to the equilibrium distribution
Eq.~(\ref{gl6.3}) and the conditions Eqs.~(\ref{gl6.9}, \ref{gl6.10}) for
the rates. We assume that the left-right antisymmetric combinations are
zero such that Eq.~(\ref{gl6.10}) is trivially satisfied. To simplify the
problem further, we choose the following solution of Eq.~(\ref{gl6.9})
\BEQ
\gamma_{1,0}^{s}(0,0) - \gamma_{1,0}^{s}(1,0) =
\gamma_{1,0}^{s}(0,1) - \gamma_{1,0}^{s}(1,1) =
\gamma_{1,1}(1,1) - \gamma_{1,1}(0,0)  =0
\EEQ
With this choice we get
\BEQ \label{wuu}
w_{1,0}(1,0) = \mu w_{1,0}(0,0) \;\; , \;\;
w_{1,0}(1,1) = \mu w_{1,0}(0,1) \;\; , \;\;
w_{1,1}(1,1) = \mu^{2} w_{1,1}(0,0)
\EEQ
This tell us that at $T = 0 (\mu = 0)$ the master equation describing the
critical dynamics of the Ising model reduces to the master equation considered
in Sec.~\ref{sect4} in which annihilation, coagulation and death processes
were considered. The particular case with only annihilation was already
noticed by Family aand Amar~\cite{Ref21}.

Taking take $w_{1,1}(1,0)=1$ and by using probability conservation we get
\BEA
w_{0,0}(0,0) &=& \mu^{2} w_{1,1}(0,0) + 2\mu w_{1,0}(0,0) \nonumber \\
w_{0,0}(1,0) &=& 1 + w_{1,0}(0,0) + \mu w_{1,0}(0,1) \nonumber \\
w_{0,0}(1,1) &=& 2 w_{1,0}(0,1) + w_{1,1}(0,0)
\EEA
Using Eqs.~(\ref{hha}), (\ref{rel4.47}) and (\ref{wuu}) we get the Hamiltonian:
\BEA
H &=& -\frac{1}{2} \sum_{i=1}^{L-1} \left[
\sig_{i}^{x} \sig_{i+1}^{x} + \sig_{i}^{y} \sig_{i+1}^{y}
+ \Delta \sig_{i}^{z} \sig_{i+1}^{z} + (1-\Delta')
\left( \sig_{i}^{z} + \sig_{i+1}^{z} \right) \right. \nonumber \\
& & + \mu w_{1,1}(0,0) \left( \sig_{i}^{x} \sig_{i+1}^{x} -
\sig_{i}^{y} \sig_{i+1}^{y} \right) \nonumber \\
& & + 2\mu^{1/2} w_{1,0}(0,0) \left( \sig_{i}^{x} E_{i+1}^{00}
+ E_{i}^{00} \sig_{i+1}^{x} \right) \nonumber \\
& & \left. + 2\mu^{1/2} w_{1,0}(0,1) \left( \sig_{i}^{x} E_{i+1}^{11}
+ E_{i}^{11} \sig_{i+1}^{x} \right) \right]
\EEA
where
\BEA
\Delta &=& 1 - \frac{1+\mu^{2}}{2} w_{1,1}(0,0) -
(1-\mu) w_{1,0}(0,1) + (1-\mu) w_{1,0}(0,0) \nonumber \\
\Delta' &=& \Delta -\mu w_{1,0}(0,1) + \mu^{2} w_{1,1}(0,0)
-(1-2\mu) w_{1,0}(0,0)
\EEA
This result is very interesting and deserves a few comments.
If $w_{1,0}(0,0) \neq 0$, even for $\mu = 0$ (when the equilibrium system
is at the critical temperature $T =0$) we still
have $\Delta' < \Delta$ and consequently the time-dependent
system is massive with an exponential fall-off in the correlation functions.
This is to the best of our knowledge the first example of a system having an
equilibrium  second order phase transition but no critical slowing down.
It is not obvious that this phenomenon generalizes into higher dimensions
since only in one dimension, at $T = T_c$ the system is fully ordered.

On the other hand if $w_{1,0}(0,0) = 0$, but $w_{1,0}(0,1)$ is
non-zero, we would expect the relaxation time should be
\BEQ
\tau^{-1} \sim \mu \label{isire}
\EEQ
The reason is simple and can be understood using perturbative arguments.
We get terms of order ${\cal O}(\mu)$ in $\Delta'$, which
couples to the $U(1)$ scalars, $\sig_{i}^{z} + \sig_{i+1}^{z}$, in
the Hamiltonian. The terms present in the Hamiltonian which are of
order ${\cal O}(\mu^{1/2})$ occur in combinations like
$\mu^{1/2} w_{1,0}(0,1)$ which do not couple to $U(1)$ scalars and
should thus only contribute in second order.
The contribution of order
${\cal O}(\mu)$ can be only eliminated when a $Z_2$ symmetry is
present in the problem. This implies that
\BEQ
w_{1,0}(0,0) = w_{1,0}(0,1) = 0
\EEQ
and by repeating the same argument we find
\BEQ \label{eq:8.8}
\tau^{-1} \sim \mu^{2}
\EEQ
However these perturbative arguments should be considered with some care since
they apply to chains of a given length L. It might happen that in the large L
limit, the leading term in the power expansions in $\mu$ has vanishing
coefficients and we have to consider the next leading order term. To be more
explicit, in a concrete calculation of the case $w_{1,0}(0,1) \neq 0$,
$w_{1,0}(0,0) = w_{1,1}(0,0) = 0$ for a given chain one obtains:
\BEQ \label{tau}
\tau^{-1}_L = a_L + b_L\mu + c_L\mu^2 + ...
\EEQ
In the large L limit $a_L$ must vanish (we end up in a massless system at
$\mu =0$ ) but the same can also happen to $b_L$. It turns out that this is
indeed happening since we can derive the exact result using the calculations of
Ref.\cite{benA90} for the coagulation-decoagulation chemical process and find
\BEQ \label{ttau}
\tau^{-1} = \frac{\mu^2}{4}.
\EEQ
We would like to emphasize that the calculation of the coefficients $a_L,
b_L, c_L, ...$ can be difficult in general so that numerical estimates might
be useful here. For the case $w_{1,1}(0,0) \neq 0$ and $w_{1,0}(0,1) =
w_{1,0}(0,0)$ the perturbative argument is correct and Eq.(\ref{eq:8.8}) stays
valid as we can see from the exact result of Ref.\cite{Ref20}. We conclude
that in the Ising case the critical exponent $z = 2$ is universal.

\section{Three-states critical dynamics} \label{sect8}

Turning back to the three-states models, we now assume, in analogy to the
two-state case, that the assymmetric rates in Eq.~(\ref{gl6.13}) are
zero and we thus only have to solve Eq.~(\ref{gl6.12}). We begin
by considering the ordinary Potts model, with $\phi=0$. We find
\BEQ
w_{2,1}(1,2) = \mu^{2} w_{2,1}(0,0) \;\; , \;\;
w_{2,1}(1,1) = \mu w_{2,1}(2,0) \;\; , \;\;
w_{2,1}(2,2) = \mu w_{2,1}(0,1)
\EEQ
This leads to the following Hamiltonian
\BEA
H &=& \sum_{i=1}^{L-1} \left[
2\mu^{2} w_{2,1}(0,0) E_{i}^{00} E_{i+1}^{00} +
2 w_{2,1}(2,0) E_{i}^{11} E_{i+1}^{11} +
2 w_{2,1}(0,1) E_{i}^{22} E_{i+1}^{22} \right. \nonumber \\
& & + \left( w_{2,1}(1,0) + \mu w_{2,1}(0,1) \right)
\left( E_{i}^{00} E_{i+1}^{11} + E_{i}^{11} E_{i+1}^{00} \right)\nonumber \\
& & + \left( w_{2,1}(0,2) + \mu w_{2,1}(2,0) \right)
\left( E_{i}^{00} E_{i+1}^{22} + E_{i}^{22} E_{i+1}^{00} \right)\nonumber\\
& & + \left( w_{2,1}(2,1) + w_{2,1}(0,0) \right)
\left( E_{i}^{11} E_{i+1}^{22} + E_{i}^{22} E_{i+1}^{11} \right)
\nonumber \\
& &-\mu w_{2,1}(0,0)\left( E_{i}^{10} E_{i+1}^{20}
+ E_{i}^{01} E_{i+1}^{02} +
E_{i}^{20} E_{i+1}^{10} + E_{i}^{02} E_{i+1}^{01}\right) \nonumber \\
& &-\mu^{1/2} w_{2,1}(2,0)
\left( E_{i}^{10} E_{i+1}^{12} + E_{i}^{01} E_{i+1}^{21} +
E_{i}^{21} E_{i+1}^{01} + E_{i}^{12} E_{i+1}^{10}\right) \nonumber \\
& &-\mu^{1/2} w_{2,1}(0,1)
\left( E_{i}^{02} E_{i+1}^{12} + E_{i}^{20} E_{i+1}^{21} +
E_{i}^{21} E_{i+1}^{20} + E_{i}^{12} E_{i+1}^{02}\right) \nonumber \\
& & - w_{2,1}(1,0) \left( E_{i}^{10} E_{i+1}^{01} + E_{i}^{01} E_{i+1}^{10}
\right) \nonumber \\
& & - w_{2,1}(2,1) \left( E_{i}^{21} E_{i+1}^{12} + E_{i}^{12} E_{i+1}^{21}
\right) \nonumber \\
& & \left. - w_{2,1}(0,2)
\left( E_{i}^{02} E_{i+1}^{20} + E_{i}^{20} E_{i+1}^{02}
\right) \right] \label{gl8.2}
\EEA
In writing Eq.~(\ref{gl8.2}), we have used the similarity transformation
(\ref{eq:2.8}) with $h^{(k)}(1)/h^{(k)}(0) = h^{(k)}(2)/h^{(k)}(0) =
\sqrt{\mu}$.
We are thus left, after choosing the time scale, with five free parameters.
If we now take $\mu=0$ in Eq.~(\ref{gl8.2}), we get back to
the chains studied in Section~\ref{sect5}. If we make the choice of coupling
constants from Eqs.~(\ref{gl5.8}, \ref{gl5.9}), at $\mu=0$ we find that
$H$ coincides with the $H_0$ of Eq.~(\ref{gl5.10}). We can thus write
\BEQ
H = H_0 + \tilde{H}_1
\EEQ
where $\tilde{H}_1$ is in the case $\alpha_{AA}=\alpha_{BB}$
\BEA
\tilde{H}_1 &=& \sum_{i=1}^{L-1} \left[
2 \mu^{2} \alpha_{AB} E_{i}^{00} E_{i+1}^{00} +
\mu \alpha_{AA} \left( ({\bf 1} - \rho^{0})_{i} \rho_{i+1}^{0}
+ \rho_{i}^{0} ({\bf 1} - \rho^{0})_{i+1} \right) \right. \nonumber \\
& & -\mu \alpha_{AB} \left( E_{i}^{10} E_{i+1}^{20} +
E_{i}^{01} E_{i+1}^{02} + E_{i}^{20} E_{i+1}^{10} + E_{i}^{02} E_{i+1}^{01}
\right) \nonumber \\
& & \left. -\mu^{1/2} \alpha_{AA} \left(
\left( E^{10}+E^{02}\right)_{i} E_{i+1}^{12} +
E_{i}^{12} \left( E^{10}+E^{02}\right)_{i+1} \right. \right. \nonumber \\
& & \left. \left. + \left( E^{01}+E^{20}\right)_{i} E_{i+1}^{21} +
E_{i}^{21} \left( E^{01}+E^{20}\right)_{i+1}
\right) \right] \label{gl8.3}
\EEA
and
\BEQ
\rho^{0} = \left( \begin{array}{ccc}
1 & 0 & 0 \\
0 & 0 & 0 \\
0 & 0 & 1 \end{array} \right)
= E^{00} + E^{22}
\EEQ
Notice that the terms which couple to $\mu \alpha_{AA}$ keep
the $U(1) \otimes U(1)$ symmetry of the unperturbed problem. If this
is the only perturbation present, we should thus get
\BEQ
\tau^{-1} \sim \mu
\EEQ
This result is similar to the behaviour observed for the Ising model (see Eq.
{}~(\ref{isire})). We would like to stress that this result might be "naive"
(see the discussion after Eq.~(\ref{eq:8.8}) in Sec.~\ref{sect7}).

We now consider the superintegrable model, where $\phi=\pi/6$.
This case is interesting because
$g_{3}(2) = 1$ and this will have important consequences. From
Eq.~(\ref{gl6.12}), we find
\BEQ
w_{2,1}(1,2) = \mu^{3} w_{2,1}(0,0) \;\; , \;\;
w_{2,1}(1,1) = \mu^{3} w_{2,1}(2,0) \;\; , \;\;
w_{2,1}(2,2) = w_{2,1}(0,1)
\EEQ
The Hamiltonian is
\BEA
H &=& \sum_{i=1}^{L-1} \left[
2\mu^{3} w_{2,1}(0,0) E_{i}^{00} E_{i+1}^{00} +
2 w_{2,1}(2,0) E_{i}^{11} E_{i+1}^{11} +
2 w_{2,1}(0,1) E_{i}^{22} E_{i+1}^{22} \right. \nonumber \\
& & + \left( w_{2,1}(1,0) + w_{2,1}(0,1) \right)
\left( E_{i}^{00} E_{i+1}^{11} + E_{i}^{11} E_{i+1}^{00} \right)\nonumber \\
& & + \left( w_{2,1}(0,2) + \mu^{3} w_{2,1}(2,0) \right)
\left( E_{i}^{00} E_{i+1}^{22} + E_{i}^{22} E_{i+1}^{00} \right)\nonumber\\
& & + \left( w_{2,1}(2,1) + w_{2,1}(0,0) \right)
\left( E_{i}^{11} E_{i+1}^{22} + E_{i}^{22} E_{i+1}^{11} \right)
\nonumber \\
& &-\mu^{3/2} w_{2,1}(0,0)
\left( E_{i}^{10} E_{i+1}^{20} +
E_{i}^{20} E_{i+1}^{10} + E_{i}^{01} E_{i+1}^{02} + E_{i}^{02} E_{i+1}^{01}
\right) \nonumber \\
& &-\mu^{3/2} w_{2,1}(2,0)
\left( E_{i}^{10} E_{i+1}^{12} + E_{i}^{01} E_{i+1}^{21} +
E_{i}^{21} E_{i+1}^{01} + E_{i}^{12} E_{i+1}^{10}\right) \nonumber \\
& &- w_{2,1}(0,1)
\left( E_{i}^{02} E_{i+1}^{12} + E_{i}^{20} E_{i+1}^{21} +
E_{i}^{21} E_{i+1}^{20} + E_{i}^{12} E_{i+1}^{02}\right) \nonumber \\
& & - w_{2,1}(1,0) \left( E_{i}^{10} E_{i+1}^{01} + E_{i}^{01} E_{i+1}^{10}
\right) \nonumber \\
& & - w_{2,1}(2,1) \left( E_{i}^{21} E_{i+1}^{12} + E_{i}^{12} E_{i+1}^{21}
\right) \nonumber \\
& & \left. - w_{2,1}(0,2)
\left( E_{i}^{02} E_{i+1}^{20} + E_{i}^{20} E_{i+1}^{02}
\right) \right] \label{gl8.6}
\EEA
Notice that for $\mu=0$ we got a new term in Eq.~(\ref{gl8.6}) which did not
exist in the Hamiltonians studied in Section~\ref{sect5} and which reads
\BEQ \label{gl8.7}
w_{2,1}(0,1) \left( \rho_{i}^{+} \sig_{i+1}^{+} + \rho_{i}^{-}
\sig_{i+1}^{-} + \sig_{i}^{-}\rho_{i+1}^{-} + \sig_{i}^{+}
\rho_{i+1}^{+} \right) \label{hqui}
\EEQ
where we have used the notation from Eq.~(\ref{notat}). It is not
known whether the Hamiltonian Eq.~(\ref{gl8.6}) for $\mu=0$ is critical
or not. Comparing the ordinary Potts and the superintegrable Potts
cases we notice that the number of independent rates
is different. We also see, by comparing Eqs.~(\ref{gl8.2}) with~(\ref{hqui})
that $\mu^{1/2}$ is replaced by $\mu^{3/2}$, consequently the dynamical
exponents changes. We conclude that  when moving from the Potts model
(symmetry $S_3$) to the superintegrable chiral Potts model (symmetry $Z_3$),
the exponents also changes.

We have seen how the universal critical exponents
can be predicted from considering
the structure of integrable Hamiltonians at $\mu=0$ and
appealling to perturbation theory. Here again,
we have implicitly assumed that all the six rates $w_{2,1}(0,0),
w_{2,1}(2,0), w_{2,1}(0,1), w_{2,1}(1,0), w_{2,1}(2,1)$ and
$w_{2,1}(0,2)$ have finite, non-va\-ni\-shing limits as $\mu\rar 0$.
Otherwise, one may create any value of some
effective exponent like in Refs. \cite{julia,forgas,lage}.

\section{Conclusions}

We have started our study asking ourselves what we thought is a
simple question: is the present progress achieved in the
understanding of one-dimensional integrable quantum chains useful for
solving master equations describing the dynamics of classical
one-dimensional spin systems ? We find plenty of evidence
for a positive answer. At same time, after finishing this long
paper, we have the feeling that we are just at the beginning of
a long path.

Although initially we thought that our task would be just to use
the available mathematical knowledge of integrable systems to
find physical results, the physical problems brought a lot of
feed-back into mathematics. Studying open chains with
particular transitions rates lead us, to our surprise,
automatically to new, hermitian and non-hermitian, representations
of interesting associative algebras. We remind the reader that in
equlibrium problems, quantum groups and associative algebras
appear only through rather artificial boundary conditions. The
physical applications of non-hermitian representations of
associative algebras appear for the first time in our context. More about
this subject will be published elsewhere.

We would like to stress once more that although the phase
diagram in the space of transition rates can be easily obtained
from the Hamiltonian spectrum computed for example by using
the Bethe ansatz, the calculation of nonequilibrium
averages (which are {\em not} those normally occuring in
equilibrium statistical physics) might pose formidable
problems.

For the physical understanding of reaction-diffusion processes
or of simple critical dynamics we think that we went beyond the
particular cases studied up to now in the literature.

As the reader has certainly noticed while going through
Sections~\ref{sect4} and \ref{sect5}, there is plenty of room
for more work especially for three-states models, where
we have only considered systems with $Z_3$ symmetry and even
in this case, the whole phase diagram is not yet completely known.

Our experience with the study of properties of quantum chains
for equlibrium purposes lead us to repeat analogous
questions for nonequlibrium problems. For example, in the
simple model studied in Section~\ref{sectA} we have
verified that finite-size scaling applies also to nonequlibrium
situations. This opens the possibility of using standard numerical
methods  of matrix diagonalization of finite systems,
for computing critical exponents. The solutions found in
Section~\ref{sect6} for the detailed-balance equations should
be useful for other dynamical processes than those discussed in
Sections~\ref{sect7} and \ref{sect8}, in a better understanding
of critical dynamics.

\zeile{3}
\noindent{\bf Acknowledgements}
\zeile{2}
We would like to thank C. G\'omez, V. Privman and G. Sch\"utz for useful
discussions. We are most grateful to Klaus Krebs, Markus Pfannm\"ulller and
Birgit Wehefritz for the excellent job they did critically reading the
manuscript.
F.A. thanks the Funda\c{c}\~{a}o de Amparo \`a Pesquisa do
Estado de S\~{a}o Paulo - FAPESP - Brasil
and the Deutsche Forschungsgemeinschaft - DFG - Germany for support.
M.D. and M.H. thank the Swiss National Science Foundation for
support.
V.R. would like to thank CERN, the University of Geneva, the Einstein
Center of the Weizmann Institute and SISSA for the warm hospitality
he has enjoyed and has enabled him to participate in this project.

\appendix       

\appsection{A}{Eigenvectors of some special non-hermitian Hamiltonians}
\label{appa}

We discuss the treatment of a certain class of non-hermitian
Hamiltonians with the structure
\BEQ
H = H_0 + H_1
\EEQ
where $H_0$ is hermitian and has the same spectrum as $H$. Moreover we are
interested in the case where the eigenspectrum is non-degenerated. Suppose
 the eigenvectors $\ket{u_i}$ and the
eigenvalues $\lmb_i$ of $H_0$
\BEQ
H_0 \ket{u_{i}} = \lambda_i \ket{u_{i}}
\EEQ
are known and we want to find the eigenvectors of $H$.
This problem is
of interest for both analytical and numerical (finite-size scaling)
calculations in reaction-diffusion processes (see Eqs. (4.11), (4.31)
and (6.11)). In the basis $\{ \ket{u_{i}} \}$, $H_1$ has only the following
non-vanishing matrix elements
\BEQ
\bra{u_{i}} H_1 \ket{u_{j}} = G_{ij} \;\; ; \;\; i<j \;\; , \;\;
i,j = 1,\ldots, N
\EEQ
Thus the eigenvalues of $H$ are again $\lambda_1,\ldots,\lambda_N$. Let
$\ket{v_{i}}$ be the eigenvector of $H$ corresponding to the eigenvalue
$\lambda_i$. It can be written as
\BEQ
\ket{v_{i}} = \sum_{i\geq j} A_{ij} \ket{u_{j}}
\EEQ
with $A_{ii}=1$ and
the graphical rule to compute the $A_{i,j} (i > j)$ is obvious if we
give the first few
\BEA
A_{21} &=& \frac{G_{12}}{\lmb_2 -\lmb_1} \;\; , \;\;
A_{32} = \frac{G_{23}}{\lmb_3 -\lmb_2} \;\; , \;\;
A_{31} = \frac{G_{13}}{\lmb_3 -\lmb_1} + \frac{G_{12}G_{23}}{(\lmb_3 -\lmb_2)
(\lmb_3 -\lmb_1)} \nonumber \\
A_{43} &=& \frac{G_{34}}{\lmb_4 -\lmb_3} \;\; , \;\;
A_{42} = \frac{G_{24}}{\lmb_4 -\lmb_2} + \frac{G_{23}G_{34}}{(\lmb_4 -\lmb_3)
(\lmb_4 -\lmb_2)} \nonumber \\
A_{41} &=& \frac{G_{14}}{\lmb_4 -\lmb_1} + \frac{G_{13}G_{34}}{(\lmb_4-\lmb_3)
(\lmb_4-\lmb_1)} + \frac{G_{12}G_{24}}{(\lmb_4-\lmb_2)
(\lmb_4-\lmb_1)} \nonumber \\
& & + \frac{G_{12}G_{23}G_{34}}{(\lmb_4-\lmb_3)
(\lmb_4-\lmb_2)(\lmb_4-\lmb_1)}
\EEA

\appsection{B}{The seven-vertex model} \label{appb}

As an example, we shall consider in more detail one of the models related to
a Hecke algebra (see Appendix C)
in order to obtain the Boltzmann weights of an
associated two-dimensional vertex model and thus prove its integrability.

The model we consider is the one introduced in Section~\ref{sect4} which
describes diffusion and pairwise annihilation of $A$ particles, see
Eq.~(\ref{eq:4.8}). We choose the special tuning which makes $\Delta=0$
and take the functions $\vph^{(k)}(\alpha,\beta)=1$ where $\alpha,\beta=0,1$.
The Hamiltonian is given by
\BEQ \label{eq:A1}
H = D( H_0 + H_1 )
\EEQ
where
\BEA
H_0 &=& - \sum_{i=1}^{L-1} \left[ q E_{i}^{01}E_{i+1}^{10} + \frac{1}{q}
E_{i}^{10}E_{i+1}^{01} +\frac{q}{2}\left( \sig_{i}^{z}-1\right)
+\frac{1}{2q}\left( \sig_{i+1}^{z}-1\right) \right] \nonumber \\
H_1 &=& - \Omega \sum_{i=1}^{L-1} E_{i}^{01} E_{i+1}^{01}
\EEA
and
\BEQ
\Omega = \frac{w_{1,1}(0,0)}{D} \;\; , \;\;
q = \sqrt{\frac{w_{1,1}(1,0)}{w_{1,1}(0,1)}} \;\; , \;\;
D = \sqrt{ w_{1,1}(0,1) w_{1,1}(1,0) }
\EEQ
Doing the canonical transformation
$E_{i}^{k\ell} =(-1)^{k-\ell} E_{i}^{k\ell}$, only at even sites $i$,
this Hamiltonian in the $\sig^{z}$ basis takes the simple form
\BEQ \label{eq:B4}
H = - D \sum_{i=1}^{L-1} e_i
\EEQ
where
\BEQ
e_i = {\bf 1}_{1} \otimes \ldots \otimes {\bf 1}_{i-1} \otimes
\left( \begin{array}{cccc}
0 & 0 & 0 & \Omega \\
0 & \frac{1}{q} & q & 0 \\
0 & \frac{1}{q} & q & 0 \\
0 & 0 & 0 & q + \frac{1}{q} \end{array}
\right) \otimes {\bf 1}_{i+2} \otimes \ldots
\EEQ
and $i=1,2,\ldots,L-1$ and ${\bf 1}_{i}$
are $2\times 2$ unit matrices attached to
the site $i$.

The Hamiltonian Eq.~(\ref{eq:B4}) is known to be
integrable through a Jordan-Wigner transformation (see Section~\ref{sectA}),
we think however that the approach given here gives
not only new insight into the
problem but is of a larger validity.

We can show that the above matrices $e_{i}, i=1,\ldots,L$ satisfy the Hecke
algebra, for arbitrary values of $\Omega$
\BEA
&& e_i e_{i\pm 1} e_i - e_i = e_{i\pm 1} e_i e_{i\pm 1} - e_{i\pm 1}
\nonumber \\
&& \left[ e_i , e_j \right] =0 \;\; ; \;\; |i-j| \geq 2 \label{eq:Hecke} \\
&& e_{i}^{2} = \left( q + q^{-1} \right) e_{i}  \nonumber
\EEA
This is the first example we know of where nonhermitian (take $q$ real)
representations of the Hecke algebra appear in physical applications. The
hermitian case $\Omega =0$ corresponds to the quantum chain introduced by
Saleur \cite{SA}.

Due to the algebraic properties Eq.~(\ref{eq:Hecke}) we can construct
an associated two-dimensional vertex model having a row-to-row
transfer matrix depending on the spectral parameter $\theta$. There
transfer matrices will satisfy the Yang-Baxter equations \cite{YB}
which implies that they commute among themselves for different values of
the spectral parameter.

In order to obtain the configuration and the Boltzmann weights associated
to this vertex model, we need the spectral parameter dependent
matrix $\vm{R}_{i} (\theta), i=1,2,\ldots$. This is found by the
Baxterization procedure \cite{Jone89} for Hecke algebras, namely
\BEQ
\vm{R}_{i} (\theta) = \frac{\sinh \theta}{\sinh \eta} e_i
+ \frac{\sinh (\eta-\theta)}{\sinh \eta} \;\; , \;\; q = e^{\eta}
\EEQ
which gives us
\BEQ
\vm{R}_{i} (\theta) = {\bf 1}_{1} \otimes \ldots \otimes {\bf 1}_{i-1}
\otimes \frac{1}{\sinh \eta} \left(
\begin{array}{cccc}
\sinh (\eta-\theta) &  &  & \Omega \sinh \theta \\
 & e^{-\theta} \sinh \eta & e^{\eta} \sinh \theta & \\
 & e^{-\eta} \sinh \theta & e^{\theta} \sinh \eta & \\
 & & & \sinh (\eta+\theta) \end{array} \right)
\otimes {\bf 1}_{i+2} \otimes \ldots
\EEQ
The relations Eq.~(\ref{eq:Hecke}) imply that these matrices satisfy
the spectral parameter dependent braid group relations
\BEA
\vm{R}_{i} (\theta) \vm{R}_{i\pm 1} (\theta+\theta') \vm{R}_{i} (\theta')
&=& \vm{R}_{i\pm 1} (\theta') \vm{R}_{i} (\theta+\theta')
\vm{R}_{i\pm 1} (\theta) \nonumber \\
\left[ \vm{R}_{i} (\theta) , \vm{R}_{j} (\theta') \right] &=& 0
\;\; , \;\; |i-j| \geq 2
\EEA
which are equivalent to the Yang-Baxter equations.

The Boltzmann  weights $S_{\ell,m}^{kn}$ of the vertex configuration labelled
by $(k,\ell,m,n)$ in the associated vertex model can be obtained from
the relation
\BEQ
\vm{R}_{i} (\theta) = S_{\ell,m}^{k,n} {\bf 1}_{1} \otimes
\ldots \otimes {\bf 1}_{i-1} \otimes E^{mk} \otimes E^{n\ell}
\otimes {\bf 1}_{i+2} \otimes \ldots
\EEQ
This implies that the vertex model associated to Eq.~(\ref{eq:A1})
is a seven vertex model. If
we denote by an index zero a down (or left) arrow
and by an index one an up (or right) arrow, the vertex configurations with
their Boltzmann weights are given in Figure~2. We also show in this figure the
corresponding chemical processes related to each vertex. The vertices 1 to 4
correspond to no reaction, vertices 5 and 6 to diffusion to the right and
the left and vertex 7 to the pair annihilation process. The derivative
of the logarithm of the row-to-row transfer matrix, with these
Boltzmann weights and evaluated at $\theta=0$, gives back the
Hamiltonian Eq.~(\ref{eq:A1}).

\appsection{C}{Hecke algebra and reaction-diffusion processes} \label{appc}

In this appendix we define the Hecke algebra and give some examples of
Hamiltonians, related to dynamical processes, which are representations
of this algebra.

The Hecke algebra is an associative algebra with generators $e_i$
($i=1,\ldots,L-1$) satisfying the relations
\BEA
&& e_i e_{i\pm 1} e_i - e_i = e_{i\pm 1} e_i e_{i\pm 1} - e_{i\pm 1}
\label{eq:C1}\\
&& \left[ e_i , e_j \right] =0 \;\; ; \;\; |i-j| \geq 2 \label{eq:C2} \\
&& e_{i}^{2} = \left( q + q^{-1} \right) e_{i}  \label{eq:C3}
\EEA
where $q$ is a complex parameter.

One of the main features of the above algebra is related to the fact that
a spectral-dependent $\vm{R}(u)$ matrix, satisfying the Yang-Baxter relations
\cite{YB}, can be constructed in a standard form. As a consequence of this
procedure, also called ``Baxterization'' \cite{Jone89}, the Hamiltonian
$\sum_{i=1}^{L-1} e_i$, as well as its associated vertex model, has an
infinite number of conservation laws and we expect, in general, its exact
integrability. In Appendix B we give an example of this ``Baxterization''
procedure and derive the associated vertex model
for one of the chains considered in this paper.

It is important to stress here that distinct Hamiltonians satisfying the
same Hecke algebra may correspond to different representations of the
algebra. They will share, apart from degeneracies (which may be zero)
the same eigenenergies.
These chains will have a massive
or massless behaviour depending on the value of $q$. In particular, for
$q$ real they will always have a massive behaviour except for $q=1$, where they
will be massless.

The $e_i$'s appearing in quantum chains quantum
chains satisfy beyond Eqs.~(\ref{eq:C1}-\ref{eq:C3})
additional relations which define quotients of the Hecke algebra.
Obviously, quantum chains obeying the same quotient have more coincidences in
their spectra, see \cite{Mart92} for details. A well-known quotient
is the Temperley-Lieb algebra \cite{TL} defined by (\ref{eq:C2},\ref{eq:C3})
and
\BEQ \label{eq:C4}
e_{i} e_{i\pm 1}e_{i} = e_{i} \;\; , \;\; e_{i\pm 1}e_{i}e_{i\pm 1}=e_{i\pm 1}
\;\; ; \;\; i=1,2,\ldots
\EEQ
A realization of this algebra is given by the XXZ chain
with surface magnetic fields
(invariant under the $U_{q}SU(2/0)$ quantum group) and the quantum Potts
chain with free ends \cite{TL,XXZ}. In Sections~\ref{sect4} and \ref{sect5}
we have used the  notation $U_{q}SU(n)$ for $U_{q}SU(n/0)$. With
the new notation $U_{q}SU(n/m)$
with $m\neq 0$ corresponds to quantum superalgebras \cite{Deg}.)
Another less known quotient
is defined by (\ref{eq:C1}-\ref{eq:C3}) and
the additional relation \cite{Mart92}
\BEQ \label{eq:C5}
\left( e_{i}e_{i+2}\right) e_{i+1} \left(q+q^{-1}-e_{i}\right)
\left(q+q^{-1}-e_{i+2}\right) =0 \;\; ; \;\; i=1,2,\ldots
\EEQ
which has a realization in the two-colors
Perk-Schultz model \cite{PS}, invariant under
the quantum group $U_{q}SU(1/1)$.

In this paper we show that several quantum chains, related to chemical
processes, are realizations of the Hecke algebra (C.1)-(C.3).
As a general outcome from
our analysis we verifyed that as long as only diffusion and interchange of
particle processes (the number of particles in each species is conserved
separately) are allowed, these realizations arise quite naturally. In the cases
where other processes are also allowed the chain in general will not
satisfy the Hecke algebra. However, as we shall see, for certain processes
and special tunings of transition rates, the corresponding chains will turn
out to satisfy the Hecke algebra again.

Although we can generalize our results to an arbitrary number of different
chemical species, following the line of this
paper we will consider here only the
cases where we have, beyond vacancies,
1 species ($A$) or 2 species ($A$ and $B$).

Let us consider initially the cases where we have only particles and
vacancies, see Section~\ref{sect4}.1. If we
only allow the diffusion process with
the transition rates satisfying Eqs.~(\ref{eq:4.9}), (\ref{eq:4.10}) we obtain
the XXZ chain with anisotropy $\Delta=(q+q^{-1})/2$
\BEQ \label{eq:C6}
{H}/{D} = H_0 = \sum_{i=1}^{L} e_i
\EEQ
where
\BEA
e_i &=& -\frac{1}{2} \left[ \sig_{i}^{x}\sig_{i+1}^{x} +
\sig_{i}^{y}\sig_{i+1}^{y} + \frac{1}{2}\left(q+\frac{1}{q}\right)
\sig_{i}^{z}\sig_{i+1}^{z} +\frac{1}{2}\left(q-\frac{1}{q}\right)
\left(\sig_{i}^{z}-\sig_{i+1}^{z}\right) \right. \nonumber \\
&& \left. -\frac{1}{2}\left(
q+\frac{1}{q}\right) \right]  \;\; ; \;\; i=1,\ldots,L-1 \label{eq:C7}
\EEA
We can verify that these
$e_i$ matrices are just a $2^L$-dimensional representation of the
Hecke algebra. Moreover, this Hamiltonian satisfies the quotient
defining the Temperley-Lieb algebra (\ref{eq:C4}). It is
important to stress that the
terms proportional to $\sig_{i}^{z}$ appear naturally in (\ref{eq:C7})
due to the diffusion mechanisms. These terms, although not present
in a periodic chain, are crucial in order to generate the
Temperley-Lieb algebra. The particular chain (\ref{eq:C7}) is
invariant under the quantum group $U_{q}SU(2/0)$.

Another example, also with only particles and vacancies, appears when beyond
diffusion we also allow the annihilation process $A+A\rar\emptyset+\emptyset$
(with rate $w_{1,1}(0,0)$). In this case, by making the special tuning
\BEQ
w_{1,1}(0,0) = w_{1,1}(0,1) + w_{1,1}(1,0)
\EEQ
and choosing the functions $\vph^{(k)}(\ell,m)=1$ we obtain
\BEQ \label{eq:C9}
H/D = \sum_{i=1}^{L-1} e_i
\EEQ
where
\BEA
e_i &=& -q E_{i}^{01}E_{i+1}^{10}
-\frac{1}{q}  E_{i}^{10}E_{i+1}^{01} - \left(q + \frac{1}{q}\right)
E_i^{01}E_{i+1}^{01}  \nonumber \\
&& + \frac{q}{2} \left( \sig_{i}^{z}-1\right)
+ \frac{1}{2q} \left( \sig_{i+1}^{z}-1\right) \;\; ; \; \; i=1,\ldots,L-1
\label{eq:C10}
\EEA
and
\BEQ \label{eq:C11}
q = \sqrt{ \frac{w_{1,1}(1,0)}{w_{1,1}(0,1)}} \;\; , \;\;
D = \sqrt{ w_{1,1}(0,1) w_{1,1}(1,0) }
\EEQ
These matrices $e_i$ satisfy the Hecke relations (C.1)-(C.3)
and are also generators of the
quotient defined by Eq.~(\ref{eq:C5}). This is the first example we
are aware of in which non-hermitian ($q$ real) representations of the Hecke
algebra appear in a physical context. In Appendix B we derive the vertex
model associated to this chain.

Let us now consider some cases where we have two types of particles ($A$ and
$B$). If we consider only the process of diffusion and interchange of
particles,
as we saw in Section~\ref{sect5}, by choosing the diffusion rates
Eq.~(\ref{eq:5.5}) we obtain the anisotropic Sutherland model \cite{Ogie87}
\BEQ
H / D = \sum_{i=1}^{L-1} e_i
\EEQ
where
\BEA
e_i &=& \frac{1}{2}\left( q+\frac{1}{q}\right) -
\left[ \sum_{a,b=0;a\neq b}^{2} E_{i}^{ab}E_{i+1}^{ba}
+\frac{1}{2}\left(q+\frac{1}{q}\right)\sum_{a=0}^{2} E_{i}^{aa}E_{i+1}^{aa}
\right. \nonumber \\
&& \left. +\frac{1}{2}\left(q-\frac{1}{q}\right)\sum_{a,b=0;a\neq b}^{2}
\mbox{\rm sign} (a-b) E_{i}^{aa}E_{i+1}^{bb} \right]
\EEA
which again satisfy the Hecke algebra. The above Hamiltonian is invariant
under the quantum group $U_{q}SU(3/0)$ (we do not give the
corresponding quotient here). Let us now consider the case where
beyond the above processes we also include the annihilation
$A+A\rar\emptyset+\emptyset$ (rate $w_{1,1}(0,0)$). If we use the
relation (6.5) between the diffusion rates and the condition
\BEQ
w_{1,1}(0,0) = \Gamma_L + \Gamma_R
\EEQ
we obtain
\BEQ
H/D' = \sum_{i=1}^{L-1} e_i
\EEQ
where
\BEA
e_i &=& -\left[ \frac{1}{q} E_{i}^{01}E_{i+1}^{10} + q E_{i}^{10}E_{i+1}^{01}
+\frac{1}{q} E_{i}^{02}E_{i+1}^{20} \right. \nonumber \\
&& \left. +q E_{i}^{20}E_{i+1}^{02}
+\frac{1}{q} E_{i}^{12}E_{i+1}^{21} +q E_{i}^{21}E_{i+1}^{12}
+\Omega E_{i}^{01}E_{i+1}^{01} \right] \nonumber \\
&&+ q\left(
E_{i}^{00}E_{i+1}^{11}+E_{i}^{00}E_{i+1}^{22}+E_{i}^{11}E_{i+1}^{22}
\right) +\frac{1}{q}\left( E_{i}^{11}E_{i+1}^{00}
+E_{i}^{22}E_{i+1}^{00}+E_{i}^{22}E_{i+1}^{11}\right) \nonumber \\
&&+\left( q+\frac{1}{q}\right) E_{i}^{11}E_{i+1}^{11} \label{eq:C16}
\EEA
and
\BEQ \label{eq:C17}
\Omega = q + \frac{1}{q} \;\; , \;\; q = \sqrt{ \Gamma_R / \Gamma_L}
\;\; , \;\; D' = \sqrt{ \Gamma_R \Gamma_L }
\EEQ
The (nonhermitian!) operators $e_i$ also satisfy the Hecke algebra. In fact
we verified that this is a property of (\ref{eq:C16})-(\ref{eq:C17})
for arbitrary values of
$\Omega$. For $\Omega=0$ the Hamiltonian reduces to the three-color
Perk-Schultz
model \cite{PS} (which has the quantum superalgebra $U_{q}SU(2/1)$ as symmetry)
and it is also related to a special point of the $t-J$ model \cite{tJ}
where exact integrability takes place \cite{BB}.

Let us return to the case where we do not have annihilation, see
Eqs.~(\ref{eq:C9},\ref{eq:C11}).
If we now forbid the process where the particles
interchange positions ($A+B\leftrightarrow B+A$)
\BEQ
w_{1,2}(1,2) = w_{2,1}(2,1) = 0
\EEQ
we obtain
\BEQ
H / D' = \sum_{i=1}^{L-1} e_i
\EEQ
with
\BEA
e_i &=& -\left[ \sum_{a=1}^{2} \left( E_{i}^{0a} E_{i+1}^{a0} +
E_{i}^{a0}E_{i+1}^{0a} \right) + q E{i}^{00}\left(E_{i+1}^{11} +
E_{i+1}^{22} \right) \right. \nonumber \\
&& - \frac{1}{q} \left. \left( E_{i}^{11} + E_{i}^{22}\right)E_{i+1}^{00}
 \right] \;\; ; \;\; i=1,\ldots,L-1
\label{eq:C20}
\EEA
This model was introduced in \cite{AKL} and the
generators satisfy the Hecke algebra.
Moreover we verified that like the $U_{q}SU(2/0)$
model (\ref{eq:C6},\ref{eq:C7})
the matrices $e_i$ are also the generators of a $3^L$-dimensional
Temperley-Lieb algebra. The symmetry of the chain (\ref{eq:C20}) is
known and described in Ref.\cite{Arnoud}.

If we now include (see Eq.~(\ref{rel5.10})) the annihilation processes
$A+B\rar\emptyset+\emptyset$,
$A+A\rar\emptyset+B$,
$B+B\rar\emptyset+A$
with the reaction rates related in the special way
\BEQ
w_{2,1}(1,0) = w_{1,2}(2,0)=w_{2,1}(2,0)=w_{1,2}(1,0) = \Gamma_L
\EEQ
\BEQ
w_{1,2}(0,1) = w_{2,1}(0,2)=w_{1,2}(0,2)=w_{2,1}(0,1) = \Gamma_R
\EEQ
\BEQ
w_{1,2}(0,0)=w_{2,1}(0,0) = \Gamma_R + \Gamma_L
\EEQ
\BEQ
q = \sqrt{\Gamma_L / \Gamma_R } \;\; , \;\; D' = \sqrt{ \Gamma_L \Gamma_R}
\EEQ
we obtain
\BEQ \label{eq:C25}
H/D' = H_0 /D' + H_1 /D' = \sum_{i=1}^{L-1} e_i \;\; , \;\;
e_i  = e_{i}^{0} + e_{i}^{1}
\EEQ
where
\BEA
e_{i}^{0} &=& \sum_{a=1}^{2} \left( q E_{i}^{00}E_{i+1}^{aa} +
\frac{1}{q}E_{i}^{aa}E_{i+1}^{00} -qE_{i}^{a0}E_{i+1}^{0a}
-\frac{1}{q} E_{i}^{0a}E_{i+1}^{a0} \right) \nonumber \\
&& + \left(q+\frac{1}{q}\right) \sum_{a,b=1}^{2} E_{i}^{aa}E_{i+1}^{bb}
\;\; ; \;\; i=1,\ldots,L-1
\EEA
and
\BEA
e_{i}^{1} &=& -\left( \frac{1}{q} E_{i}^{01}E_{i+1}^{21}
+q E_{i}^{21}E_{i+1}^{01} + q E_{i}^{12}E_{i+1}^{02}
+\frac{1}{q} E_{i}^{02}E_{i+1}^{12} \right. \nonumber \\
&& \left. +\left(q+\frac{1}{q}\right)E_{i}^{01} E_{i+1}^{02}
+ \left( q+\frac{1}{q}\right)
E_{i}^{02}E_{i+1}^{01}\right)
\EEA
The Hamiltonian $H_0 /D' = \sum_{i=1}^{L-1} e_{i}^{0}$ was
introduced in \cite{AKL}, (see also Ref. \cite{Gome92} for the case $q = 1$)
 as a $3^L$-dimensional representation of the
Hecke algebra. We verified, guided by the physical processes of
diffusion and annihilation, the the operators $e_i$ ($i=1,\ldots,L-1$)
also satisfy
the same algebra. Morevover, we checked that both $e_{i}^{0}$ and
$e_i$ satisfy the quotient (\ref{eq:C5}) like the $U_{q}SU(1/1)$ chains,
for the underlying quantum symmetry of
(\ref{eq:C25}) see Ref.\cite{Arnoud}.

\zeile{1}\noindent{\bf Figure captions}
\zeile{3}\noindent
Figure~1: Phase diagram of the XXZ quantum chain with anisotropy $\Delta$
in a magnetic field $h$. $I$ is the massive (frozen)
ferromagnetic phase, $II$ the
massless phase which is commensurate for $h=0$ and incommensurate for
$h\neq 0$ and $III$ is the massive (frozen) antiferromagntic phase. The line
$\Delta=1-h$ corresponds to a Pokrovsky-Talapov transition.
\zeile{2}\noindent
Figure~2: Boltzmann weights of the two-dimensional vertex model associated to
the Hamiltonian Eq.~(\ref{eq:A1}). The first column gives
the weights, the second
one the arrow diagram and the third one the relationship between the
vertices and the elementary reaction-diffusion processes.


\end{document}